\documentclass[aps,twocolumn,showpacs,pra,superscriptaddress,floatfix]{revtex4}

\usepackage{amsmath,amssymb,bm} 
\usepackage{graphicx,xcolor}    

\DeclareMathOperator{\Tr}{Tr}         

\newcommand{\al}{\alpha}            
\newcommand{\bt}{\beta}             
\newcommand{\dl}{\delta}            
\newcommand{\ga}{\gamma}            
\newcommand{\ka}{\kappa}           
\newcommand{\ox}{\otimes}           
\newcommand{\vs}{\varsigma}         

\newcommand{\bra}[1]{\langle#1|}    
\newcommand{\D}{\mathcal{D}}        
\newcommand{\dn}{{\mathord{\downarrow}}} 
\newcommand{\half}{\tfrac{1}{2}}    
\newcommand{\ket}[1]{|#1\rangle}    
\newcommand{\R}{\mathbb{R}}         

\newcommand{\up}{{\mathord{\uparrow}}} 
\newcommand{\word}[1]{\quad\mbox{#1}\quad} 
\newcommand{\x}{\times}             
\newcommand{\7}{\dagger}            

\renewcommand{\H}{\mathrm{H}}       
\renewcommand{\H}{\mathcal{H}}      

\newcommand{\vecform}{\bm}              
\newcommand{\pp}{\vecform{p}}           
\newcommand{\rr}{\vecform{r}}           
\newcommand{\RR}{\vecform{R}}           
\newcommand{\xx}{\vecform{x}}           

\begin{document}

\title{Quasipinning and selection rules for excitations in atoms and molecules}

\author{Carlos L. Benavides-Riveros}
\affiliation{Departamento de F\'isica Te\'orica, Universidad de
Zaragoza, 50009 Zaragoza, Spain}
\affiliation{Instituto de Biocomputaci\'on y F\'isica de
Sistemas Complejos, 50018 Zaragoza, Spain}
\affiliation{Physikalische und Theoretische Chemie,
Universit\"at des Saarlandes, 66123 Saarbr\"ucken, Germany}

\author{Michael Springborg}
\affiliation{Physikalische und Theoretische Chemie,
Universit\"at des Saarlandes, 66123 Saarbr\"ucken, Germany}
\affiliation{School of Materials Science and Engineering, Tianjin University, Tianjin, 300072,
People's Republic of China}

\date{\today}

\begin{abstract}
Postulated by Pauli to explain the electronic structure of atoms and
molecules, the exclusion principle establishes an upper bound of~1 for
the fermionic natural occupation numbers~$\{n_i\}$. A recent analysis
of the pure $N$-representability problem provides a wide set of
inequalities for the $\{n_i\}$, leading to constraints on these
numbers. This has a strong potential impact on reduced density matrix
functional theory as we know it. In this work we continue our study
the nature of these inequalities for some atomic and molecular
systems. The results indicate that (quasi)saturation of some of them
leads to selection rules for the dominant configurations in
configuration interaction expansions, in favorable cases providing
means for significantly reducing their computational requirements.
\end{abstract}



\maketitle 

\section{Introduction}

The fermionic natural occupation numbers (arranged in the customary
decreasing order $n_i \ge n_{i+1}$) fulfill the constraint $n_i\leq1$,
allowing no more than one electron in each quantum state. This
condition, formulated by Coleman \cite{Coleman}, is necessary and
sufficient for a one-body reduced density matrix (1-RDM) to be the
contraction of an \textit{ensemble} $N$-body density matrix, provided
that $\sum_in_i= N$.

In a seminal work, Borland and Dennis \cite{losprecursores} observed
that for the rank-six approximation of a \textit{pure-state} $N=3$
system, belonging to the Hilbert space $\wedge^3\H_6$, the occupation
numbers satisfy the following additional conditions: $n_{1} + n_{6} =
n_2 + n_5 = n_3 + n_4 = 1, \; n_4 \le n_5 + n_6$. The set of
equalities allows exactly \textit{one} electron in the natural
orbitals $r$ and $7-r$. The analysis by Klyachko and coworkers
\cite{Klyachko, Alturulato} of the \textit{pure} $N$-representability
problem for the 1-RDM establishes a systematical approach,
generalizing this type of constraints. For a pure quantum system of
$N$ electrons arranged in $m$ spin orbitals, the occupation numbers
satisfy a set of linear inequalities, known as \textit{generalized
Pauli constraints} (GPC),
\begin{align}
D^\mu_{N,m}(\vecform{n}) = \ka^\mu_0 + \ka^\mu_1 n_1 
+ \cdots + \ka^\mu_m n_m \geq 0,
\label{eq:ineq}
\end{align}
with $\vecform{n} := (n_1,\dots,n_m)$, the coefficients $\ka^\mu_j \in
\mathbb{Z}$ and $\mu = 1, 2, \dots, r_{N,m}$. These conditions define
a convex polytope of allowed states in $\R^m$. When one
of the GPC is completely saturated [i.e., the equality holds in
Eq.~\eqref{eq:ineq}], the system is said to be \textit{pinned}, and it
lies on one of the facets of the po\-ly\-to\-pe.

The nature of those conditions has been explored till now in a few
systems: a model of three spinless fermions confined to a
one-dimensional harmonic potential \cite{ETH}, the lithium
isoelectronic series \cite{Sybilla}, and ground and excited states of
some three- and four-electron molecules for the rank being equal to
twice the number of electrons~\cite{Mazziotti}. For reasons that
remain mysterious, for all these systems some
inequalities are (quite often) \textit{nearly saturated}, that is, in
equations like~\eqref{eq:ineq} equality almost holds~\cite{Oxford}. 
This is the
so-called \textit{quasipinning} phenomenon, originally proposed by
Schilling, Gross and Christandl~\cite{ETH}.

The GPC force a promissory rethinking of reduced density matrix
functional theory, with possibly revolutionary
consequences~\cite{Halle}. Also, violation of the GPC has
 recently been identified as an encoder of the openness of a 
 quantum system \cite{Mazziotti2}. 

Since the dimension of the Hilbert space in the
configuration interaction (CI) method grows binomially with the number
of electrons and of spin orbitals of the system, the method easily
becomes very demanding numerically. Moreover, the CI expansion
typically contains a great number of configurations that are
superfluous (with very small expansion coefficients) for computing
molecular electronic properties. Several approaches have been devised
for selecting the most effective configurations in CI expansions
\cite{Ivanic, Cafarel}. Quasipinning offers another alley towards this
end.

Let us consider one of the conditions of Eq.~\eqref{eq:ineq}, 
$\mu$, for which pinning
\begin{align}
D^\mu_{N,m}(\vecform{n})=0
\label{eq:pinned}
\end{align}
holds. An important \textit{super-selection rule} emerges for pinned
wave functions \cite{Klyachko2}. In fact, given a pinned system
satisfying equality \eqref{eq:pinned}, the corresponding wave function
is an ei\-gen\-function of a certain operator with eigenvalue~zero. As
it will be discussed in this paper, pinning enables the wave function
to be described by an \textit{Ansatz} based on this selection rule,
reducing the number of Slater determinants in the CI expansion.
Recently, the stability of this selection rule (the potential loss of
information when assuming pinning instead of quasipinning) has been
measured for systems with non-degenerated natural orbitals which are
close to the boundary of the polytope \cite{Schilling}.

Here we examine the connection between pinning, quasipinning and the
excitation structure of the CI wave function in more detail. We
identify those configurations that are negligible when imposing
pinning on the wave function, and study the issue of the robustness of
quasipinning with increasing rank.

The paper is organized as follows. Section~\ref{super} elucida\-tes the
super-selection rule for pinned systems. Section~\ref{spin-restricted}
is of mathematical nature, as well. It is shown there is still new
wine in the old Borland--Dennis bottles: we prove, for not very
strongly correlated systems, that the spin-compensated open-shell
system $\wedge^3\H_6$ is \textit{always} pinned to the boundary of the
polytope described by the Borland--Dennis conditions. We then unveil a
new regime for spin-compensated, strongly correlated systems, and
finally we briefly discuss the relation of GPC to the linear
equalities for reduced density matrices analyzed by Davidson and
coworkers over many years~\cite{DavidsonXX,DavidsonXXI}.

In Sections~\ref{sec:lithium} and~\ref{sec:molecular} we present 
results of numerical investigations for
some atomic and molecular models: respectively a lithium atom with
broken spherical symmetry and the three-electron molecule He$^+_2$. In
Section~\ref{excitations3}
we explore the connections between quasipinning, pinning and the
excitation structure of the CI wave function for three-electron
systems. In the following section we discuss four-electron systems.
Finally, in the last Section~\ref{conclusion} we summarize our conclusions.
Throughout the paper we employ Hartree's atomic units.


\section{Super-selection rules}
\label{super}

In the full CI picture, the wave function in a given one-electron
basis is expressed as a linear combination of all possible Slater
determinants:
\begin{align}
\ket{\Psi} = \sum_K c_K \ket{\mathbf{K}}, 
\label{eq:CI}
\end{align}
with $\ket{\mathbf{K}}$ denoting a determinant. Whenever we write
expressions of this type in this paper, they are eigenfunctions of the 
spin operator
$\mathbf{S}_z$, belonging to the same eigenvalue. In general, they will
not be eigenfunctions of $\mathbf{S}^2$, so a 
spin adaptation is needed~\cite{Pauncz}.

A one-body density matrix is com\-pa\-tible with the pure-state
density matrix $\ket{\Psi}\bra{\Psi}$ whenever its spectrum satisfies
a set of linear inequalities of the type \eqref{eq:ineq}. For pinned
systems, such that the condition~\eqref{eq:pinned}~holds, the
corresponding wave function belongs to the 0-eigen\-space of the
operator
\begin{align*}
\mathbf{D}^\mu_{N,m} = \ka^\mu_0 \mathbf{1}+ \ka^\mu_1 a^\7_1 a_1 
+ \cdots + \ka^\mu_m a^\7_m a_m ,
\end{align*}
where $a^\7_i$ and $a_i$ are the fermionic creation and annihilation
operators of the state $i$. By using the expression of the wave
function in the full CI picture, this condition can be recast into a
\textit{super-selection rule} for the Slater determinants that appear
in the CI decomposition. Given a pinned system that satisfies
equality~\eqref{eq:pinned}, each Slater determinant appearing in the
expansion~\eqref{eq:CI} must be an eigenfunction of
$\mathbf{D}^\mu_{N,m}$ with an eigenvalue equal to zero. The
superfluous or ineffective configurations are thus iden\-tified by
means of the criterion \cite{Klyachko2}
\begin{equation*}
\text{if } \mathbf{D}^\mu_{N,m} \ket{\bm{K}} \neq 0, \text{ then } c_K
= 0.
\end{equation*}
This latter statement, for non-degenerate occupation numbers, follows
from a relatively well known result in symplectic geometry, whose
proof can be traced back to the eighties~\cite{Walter}. The degenerate
case needs a different kind of proof, which is forthcoming
\cite{GrossetlopesetCh}. It immediately demonstrates that the 
(quasi)pinning phenomenon allows one to drastically reduce the number of
Slater determinants in CI expansions.

The criterion becomes even more strict when more than one pinning
constraint is satisfied. Were for a set of constraints $\{\mu_1,
\mu_2,\dots, \mu_r\}$ all the GPC to saturate, the ineffective
configurations would satisfy
$$
\word{if $\mathbf{D}^{\mu_1}_{N,m}\mathbf{D}^{\mu_2}_{N,m}\cdots
\mathbf{D}^{\mu_r}_{N,m} \ket{\mathbf{K}} \neq 0$, then $c_K = 0$.}
$$
Notice that the order of the operators $\mathbf{D}^\mu_{N,m},\,
\mathbf{D}^\nu_{N,m}$ is irrelevant, since they commute.

In the remaining sections of this paper, among other things we explore
(in)effective configurations when a certain number of pinning
conditions are imposed. We mainly deal with three-electron systems,
with Hilbert space $\wedge^3\H_m,\,m\ge6$.

\section{Exact pinning in spin- compensated configurations 
for $\wedge^3\H_6$}
\label{spin-restricted}

\begin{figure}[t] 
 \centering
\includegraphics[width=6.5cm]{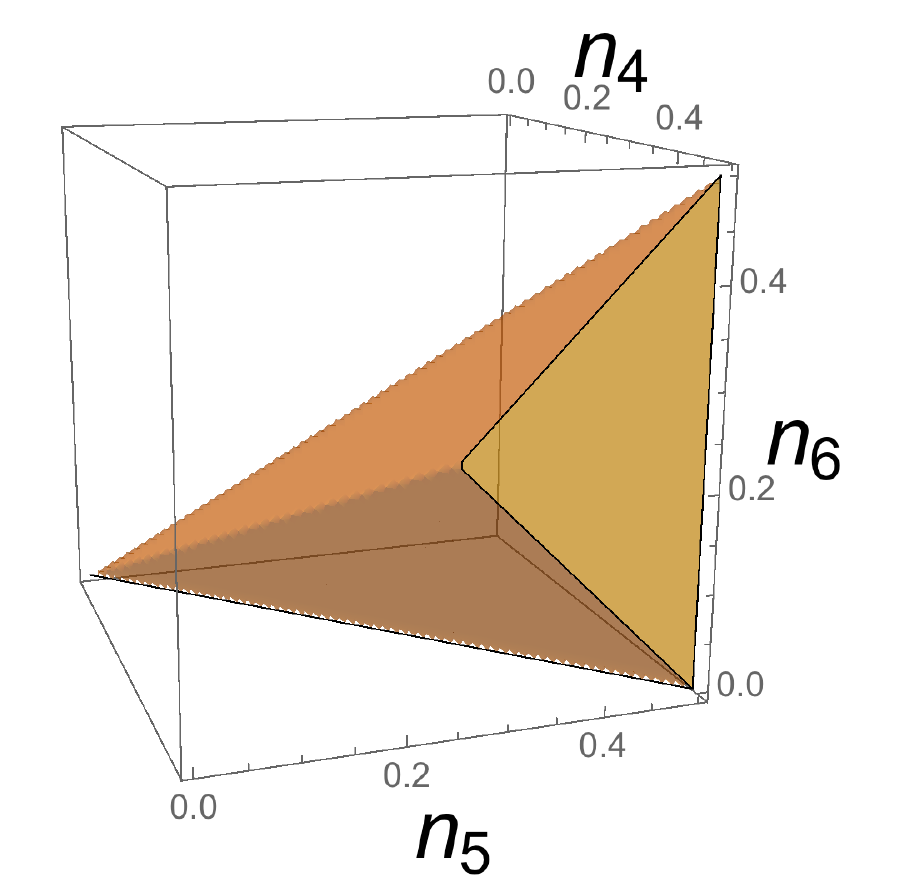}
\caption{Polytope defined by the expression $n_4 \le n_5 + n_6$,
subject to the condition $0 \le n_6 \le n_5 \le n_4 \le 0.5$. The
saturation condition $n_5 + n_6 = n_4$ is satisfied by the points
 on one of the faces of the polytope, whereas on the edges $n_5 + n_6 = 0.5$
 we have $n_4 = 0.5$ and for $n_4 = n_5$ we have $n_6 = 0$. The
 single determinant state is placed at the lower left corner $n_i =0$
 of the polytope.}
 \label{graf:polytope}
 \end{figure}

For the rank-six approximation for three-electron systems it is known
\cite{losprecursores} that the na\-tu\-ral occupation numbers satisfy
the constraints $n_r + n_{7-r}=1$ ($r = 1, 2, 3$) and
\begin{align}
2- n_1 - n_2 - n_4 \geq 0,
\label{eq:B&D}
\end{align}
where the numbers $\{n_i\}$ are arranged in the customary decreasing
order $n_{i} \geq n_{i+1}$ and fulfill the Pauli condition $n_1 \leq
1$. The inequality~\eqref{eq:B&D} together with the decreasing
ordering rule define a polytope~(Fig.~\ref{graf:polytope}). 
Clearly, the smallest possible value for the first three
occupation numbers and largest for the three last is 0.5.

Conditions $n_r+n_{7-r}=1$ imply that in the natural orbital basis,
namely $\{\al_i\}_{i=1}^6$, every Slater determinant is composed of
three natural orbitals $\ket{\al_i\al_j\al_k}$, each one belonging to
one of three different sets, say
$$
\al_i \in \{\al_1,\al_6\}, \quad  \al_j \in \{\al_2,\al_5\} \word{and}
\al_k \in \{\al_3,\al_4\}.
$$ 
This results in eight possible configurations,
\begin{align*}
\ket{\al_1\al_2\al_3},\, \ket{\al_1\al_2\al_4},\, \ket{\al_1\al_3\al_5}, \,
\ket{\al_1\al_4\al_5} , \\
\ket{\al_2\al_3\al_6},\, \ket{\al_2\al_4\al_6}, \,
\ket{\al_3\al_5\al_6},\, \ket{\al_4\al_5\al_6}.
\end{align*}

A spin-compensated configuration consists of three spin orbitals whose
spin points down, and the other three point up. Such a configuration
is in general favorable for the energy in comparison with other types
of arrangements \cite{Sybilla}. The 1-RDM (a $6\x6$ matrix) is the
direct sum of two ($3\x3$) matrices, one related to the spin up and
the other related to the spin down:
$$
\rho_1 = \rho_1^\up \oplus \rho_1^\dn.
$$ 
The wave function is an eigenstate of the total spin operator
$\mathbf{S}_z$ (and of $\mathbf{S}^2$ in the spin-restricted case).
Therefore, each acceptable Slater determinant will contain two spin
orbitals pointing up (for instance) and one pointing down. It follows
that the trace of one of those matrices will be equal to one, while
the sum of the diagonal elements of the other one will be equal to
two. Say, 
$$
\Tr \rho_1^{\up} = 2 \word{and} \Tr \rho_1^{\dn} = 1.
$$

For not very strongly correlated systems, two of the first three
occupation numbers belong to the matrix whose trace is equal to two.
Hence, we have the following two conditions: $n_i + n_j + n_x = 2$ and
$n_k + n_y + n_z = 1$, where $i,j,k \in \{1,2,3\}$ and $x,y,z \in
\{4,5,6\}$. For a given $i$ and $j$ there are three possible values of
$x$ and therefore there are in principle nine possible solutions,
\begin{align*}
n_1 + n_2 + n_x = 2 \word{and}  n_3 + n_y + n_z = 1, 
\\
n_1 + n_3 + n_x = 2 \word{and}  n_2 + n_y + n_z = 1,
\\
n_2 + n_3 + n_x = 2 \word{and} n_1 + n_y + n_z = 1 .
\end{align*}
However, we may easily dismiss all but one of them.
For instance, the case
\begin{align*}
n_2 + n_3 + n_6 = 2 \word{and}  n_1 + n_4 + n_5 = 1
\end{align*}
is impossible: using $n_1 = 1- n_6$, one obtains $-n_6 + n_4 + n_5 =
0$. This would imply that $n_4 = n_6 - n_5 \le 0$, which is out of
question. Also, for
\begin{align*}
n_1 + n_3 + n_5 = 2 \word{and}  n_2 + n_4 + n_6 = 1, 
\end{align*}
using that $n_2 = 1- n_5$, one obtains $-n_5 + n_4 + n_6 = 0$, which
would imply that $n_4 = n_5 - n_6 < n_5$. Other cases are easily seen
to give rise to rank at most four or five for the wave function,
except for
\begin{align*}
n_1 + n_2 + n_4 = 2 \word{and}  n_3 + n_5 + n_6 = 1,
\end{align*}
which saturates the re\-pre\-sentability condition~\eqref{eq:B&D}.
Therefore, for not very strongly correlated systems the
spin-compensated wave function of $\wedge^3\H_6$ lies on one of the
facets of the Borland--Dennis--Klyachko polytope. This is in agreement
with the numerical results obtained previously \cite{Sybilla,
Mazziotti}.

The wave function for this configuration for
$\wedge^3\H_6$ in the basis of natural orbitals can now be written in
terms of the 1-RDM matrix:
\begin{align}
\ket{\Psi}_{3,6} &= c_{123} \ket{\al_1\al_2\al_3} + c_{145}
\ket{\al_1\al_4\al_5} \nonumber \\ &\quad + c_{246}
\ket{\al_2\al_4\al_6};
\label{eq:ranksix} 
\end{align}
with the proviso $|c_{123}| \geq |c_{145}| \geq |c_{246}|$. It is now
patent that $\ket{\Psi}$ can be elegantly rewritten  as
\begin{align}
\sqrt{n_3} \, \ket{\al_1\al_2\al_3} + \sqrt{n_5} \,
\ket{\al_1\al_4\al_5} + \sqrt{n_6} \, \ket{\al_2\al_4\al_6},
\label{eq:yupi!}
\end{align}
in analogy to the L\"owdin--Shull (LS) functional for the two-electron
case~\cite{LS}. Note that, just like in the LS~functional, only doubly
excited configurations are here permitted~\cite{Ccomment}. (We
understand excitations with respect to the ``best density'' Slater
determinant, in the sense of~\cite{KS68}.)

The pinned configuration
$$
(n_1,n_2,n_3) = \bigl(\tfrac34,\tfrac34,\tfrac12\bigr)
$$
is far from the ``Hartree-Fock'' $(1,1,1)$ state. Now, a little
surprise awaits us: for spin-compensated, very strongly correlated
systems it is possible to show by the same method as above, the
following identity
\begin{align}
n_1 + n_2 + n_3 &= 2,
\nonumber \\
\word{equivalently} n_4 + n_5 + n_6 &= 1.
\label{eq:newpinning}
\end{align}
In terms of $\rho_1$ the wave function then reads 
\begin{align*}
\ket{\Psi} = \sqrt{n_4} \ket{\al_1\al_2\al_4} +\sqrt{n_5}
\ket{\al_1\al_3\al_5} + \sqrt{n_6} \ket{\al_2\al_3\al_6},
\end{align*}
living in the $0$-eigenspace of the operator
$$
\mathbf{2} -  a^\7_1 a_1 - a^\7_2 a_2 -  a^\7_3 a_3.
$$
We note that overlap of those wave functions with the
$\ket{\al_1\al_2\al_3}$ state is \textit{zero}. For the case
$n_4=n_5=n_6=1/3$ this was already noted by Kutzelnigg and
Smith~\cite{KS68}. The Borland--Dennis--Klyachko constraint becomes
 in~this ca\-se:
$$
2 - (n_1 + n_2 + n_4) = n_3 - n_4 \ge 0.
$$
Therefore in this regime it is possible to determine $\ket{\Psi}$ from
$\rho_1$ even without Klyachko pinning. The border between the two
regimes is given by the degeneracy line $n_3 = n_4 = \half$.
Inequality~\eqref{eq:B&D} together with the
pinning~\eqref{eq:newpinning} cut out a new facet on the polytope of
allowed states (Fig.~2).

 \begin{figure}[t] 
 \centering
  \includegraphics[width=7cm]{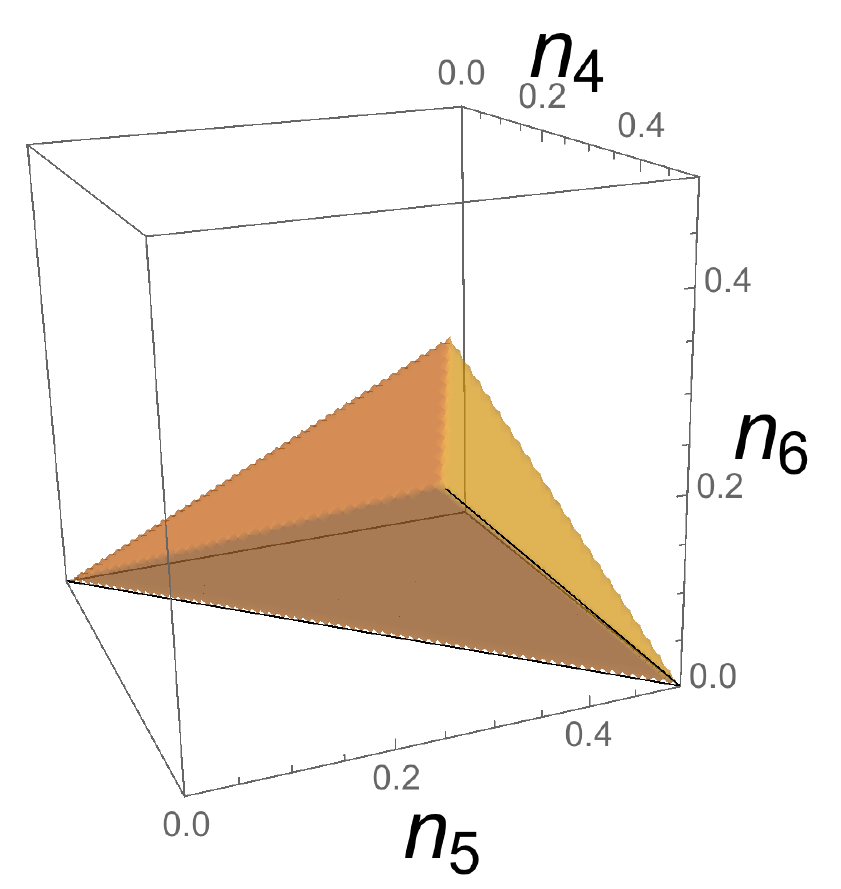} 
 \caption{Polytope defined by the expression $n_4 \le n_5 + n_6$,
subject to the condition $0 \le n_6 \le n_5 \le n_4 \le 0.5$, 
plus the condition $n_4 + n_5 + n_6 \le 1$. 
 The saturation condition $n_4 + n_5 + n_6 =1$ is satisfied
 by the points on the face of the polytope whose 
 vertices are $\bigl(\tfrac12,\tfrac12,0\bigr)$, 
 $\bigl(\tfrac12,\tfrac14,\tfrac14\bigr)$ and 
 $\bigl(\tfrac13,\tfrac13,\tfrac13\bigr)$.}
 \label{graf:polytope1}
 \end{figure}

In summary, the Borland--Dennis--Klyachko polytope of states is still
too large. In fact, the spin-compensated sta\-tes lie either on the
$n_1 + n_2 + n_4 = 2$ facet of the polytope (when closer to the
single-determinant state) or on the plane $n_4 + n_5 + n_6 =1$ (when
farther from the single-determinant state). The edge $n_3 = n_4$ is
shared by these two planes. Since the exact expressions given above
for the spin-compensated formulation of the system $\wedge^3\H_6$ lead
to a diagonal 1-RDM, without any restriction on the amplitudes
$c_{ijk}$ (provided, of course, that the orbitals are orthonormal),
for such a simple system one does not need a previous CI calculation to
compute the natural orbitals and the value of the ground-state energy 
\cite{Oxford}.

All of the 70 pure state Slater hull inequalities, grouped in four
permutation classes, shown in~\cite{DavidsonXX,DavidsonXXI} for
expectation values of products of number operators can be derived
trivially from the GPC by Klyachko; as an example:
\begin{align*}
&\langle(a^\7_aa_a + a^\7_ba_b)(a^\7_ca_c + a^\7_da_d) +
a^\7_aa_aa^\7_ba_b
\\
&\quad + a^\7_ca_ca^\7_da_d + a^\7_ea_ea^\7_fa_f\rangle \ge 1, \quad 1
\le a,b,c,d,e,f \le 6.
\end{align*}
Some are pinned. Lack of space prevents us from going into the details
of this.

The converse statement looks unlikely to us.

\section{Numerical investigations: lithium with broken spherical symmetry}
\label{sec:lithium}

In the previous paper \cite{Sybilla} we obtained rank-six, -seven and
-eight approximations for the lithium isoelectronic series by using a
set of helium-like one-particle wave functions in addition to one
hydrogen-like wave function. Guided by the classical work of Shull and
L\"owdin~\cite{LS}, for the former we employed the following set of
orthonormal spatial orbitals:
$$
\dl_n(\al, \rr) := D_n\sqrt{\frac{\al^3}{\pi}}L^2_{n-1}\bigl(2\al
r\bigr)e^{-\al r}, \quad n = 1,2,\ldots
$$
where $D^{-2}_n=\binom{n-1}{2}$, and we use the standard definition of
the associated Laguerre polynomials $L^\zeta_n$
\cite{reliablerussian}. For the hydrogen-like function we used
$$
\psi(\bt,\rr) = \frac14\sqrt{\frac{\bt^5}{6\pi}}r \,
e^{-\bt r/2}.
$$
Applying a variational procedure for the state $\ket{\dl_1 \up \dl_1
\dn \psi \up}$ results in $\al = 2.68$ and $\bt = 1.27$, and the total
energy associated to this Slater determinant becomes $-7.4179$~a.u.,
reasonably close to the Hartree--Fock energy $-7.4327$~a.u.

Now we examine the GPC when the spherical symmetry of the central
potential is broken by considering the following Hamiltonian:
\begin{align}
&H(Z,\ga) = \frac12 \sum^3_{i=1}|\pp_i|^2 
\label{eq:Hamiltonian}
\\
&\quad - \sum^3_{i=1} \frac{Z}{|\rr_i|}\bigg(1 + \ga
\frac{x^2_i}{|\rr_i|^2}\bigg) + \sum^3_{i\,<\,j}
\frac1{|\rr_i-\rr_j|}.
\nonumber
\end{align}
The case $H(3,0)$ is the Hamiltonian of lithium, whose accurate 
energy value is $-7.47806$~a.u.

A motivation behind this model is to lift constraints on the possible
occupation numbers due to the spherical symmetry of the isolated
Li~atom. Lowering the symmetry makes the model more flexible and
allows to envisage more general cases. In addition, the model can
serve to describe a Li atom embedded into some environment that does
not provide covalent interactions with the Li~atom. We have performed
the calculations of this section by searching those values of $\al$
and $\bt$ for which the approximation to the ground state leads to the
minimum energy with spin-compensated linear combinations of Slater
determinants. Analytical expressions for the electron integrals were
computed using Mathematica \cite{Mth} and orthonormalized orbitals
were obtained by the Gram--Schmidt procedure. Computations were
per\-formed with 36 decimals floating-point precision.

\subsection{Rank six}

The spin-restricted rank-six approximation for $H(3,\ga)$ 
is always pinned to the boundary of the polytope, as we
already have shown in the general case in
Section~\ref{spin-restricted}. It is interesting to examine the
spectral trajectory of the ``best'' spin-restricted state in
$\wedge^3\H_6$ as a function of the parameter $\ga$ by means of
minimizing the CI states on the manifold $(\al,\bt)$. To this end we
choose as a one-particle Hilbert space the set
$$
\{\dl_1 \up, \dl_1 \dn, \psi
\up,\psi \dn,\dl_2 \up, \dl_2 \dn\}.
$$
The Hilbert space factorizes then in the direct product of two
spin-orbital sectors $\wedge^3\,\H_6 = \H_3 \ox \wedge^2\H_3$. There
are~nine configurations in all, eight of which belong to the $j =
\half$ representation,
\begin{align*}
&\ket{\dl_1\up\dl_1\dn\psi \up}, \, \ket{\dl_1\up\dl_1\dn\dl_2 \up}, \,
\ket{\psi\up\psi\dn\dl_1 \up}, \, \ket{\psi\up\psi
\dn\dl_2\up}, \\ &\ket{\dl_2\up\dl_2\dn\dl_1 \up}, \,
\ket{\dl_2\up\dl_2\dn\psi \up}, \,
\ket{\dl_1\up\psi\dn\dl_2 \up} - \ket{\dl_1\dn\psi\up\dl_2 \up}, \\&
\ket{\dl_1\up\psi\up\dl_2 \dn} - \ket{\dl_1\dn\psi\up\dl_2 \up}.
\end{align*}
(The configuration $\ket{\dl_1\up\psi\dn\dl_2 \up}+
\ket{\dl_1\dn\psi\up\dl_2 \up}+\ket{\dl_1\up\psi\up\dl_2 \dn}$ belongs
to the representation~$j=\tfrac32$.)

 \begin{figure}[!t]
 \centering 
 \includegraphics[width=6cm]{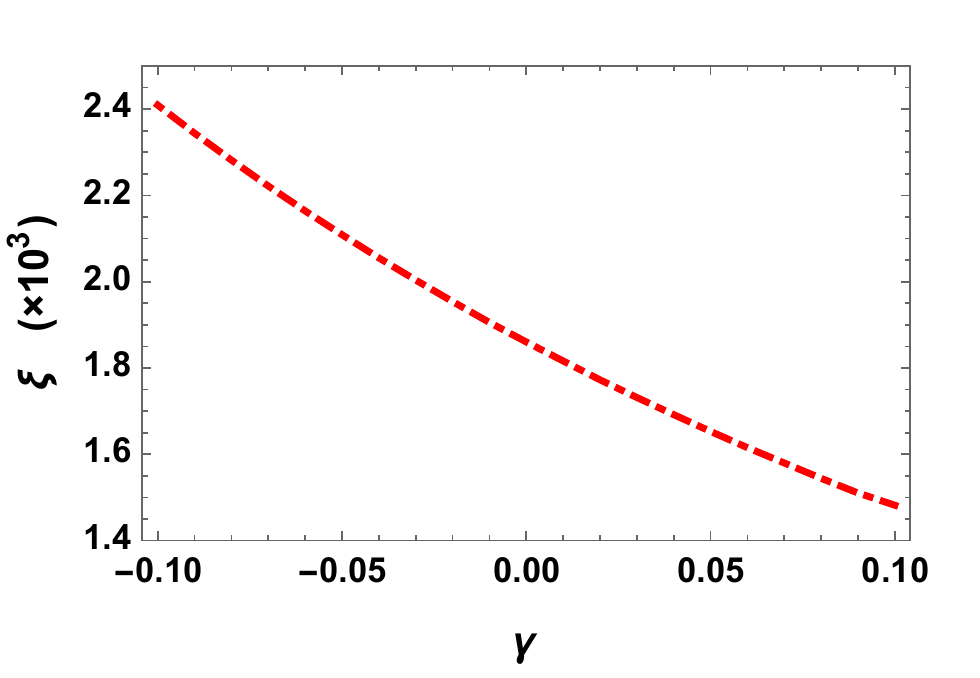}
 \caption{The distance $\xi$ between the spectrum of the ground state to the 
 extreme point of the polytope as a function of $\ga \in [-0.1,0.1]$ for the 
 spin-restricted rank-six approximation to the Hamiltonian $H(3,\ga)$ given 
 by \eqref{eq:Hamiltonian}.}
 \label{figure1}
 \end{figure}

In order to quantify the position of the set of occupation numbers on
the boundary of the polytope, one may define $\xi^2:=\sum_{i=1}^3
(1 - n_i)^2$ as the euclidean distance between the spectrum of the state to the
extreme point $(1,1,1)$ of the polytope, corresponding to the spectrum
of a single determinant. Fig.~\ref{figure1} shows $\xi$ for small values of
$\gamma$ of the electronic Hamiltonian given in~\eqref{eq:Hamiltonian}. For
decreasing $\ga$ the spectrum of the one-body density departs further
from that extremum. The kinematics however keeps the state pinned to
the boundary of the Borland--Dennis--Klyahcko polytope, since its
natural occupation numbers maintain the condition $1+ n_3 = n_1 +
n_2$.

A similar behavior is observed for a unrestricted description
using, for instance, the set
$$
\{\dl_1 \up, \dl_1 \dn, \psi \up,\dl_2 \up, \dl_2 \dn, \dl_3 \up\}
$$
as the one-particle Hilbert space. However, the energy predicted by
this latter configuration is slightly worse than the one predicted by
the spin-restricted case.

 \begin{figure}[!t]
 \centering 
 \includegraphics[width=6.2cm]{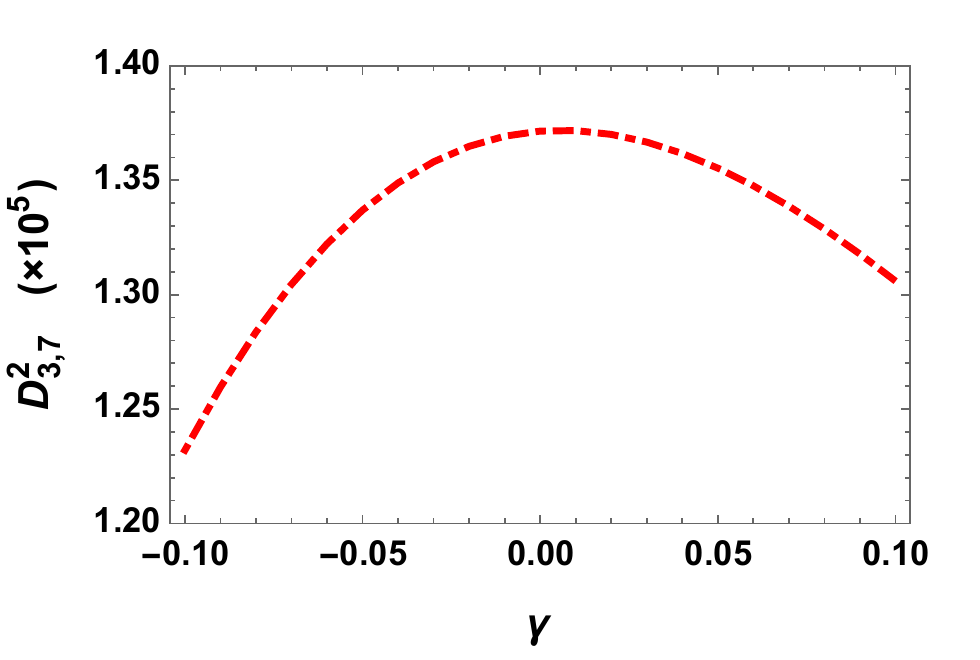}
 \caption{Second GPC $D^2_{3,7}$ for the ground-state of 
 the Hamiltonian $H(3,\ga)$ in a rank-seven approximation as a function 
 of $\ga \in [-0.1,0.1]$. For $\ga = 0.01$, the constraint reaches its maximum 
 value (namely, $1.3717 \x 10^{-5}$).}
 \label{figure4}
 \end{figure}

\subsection{Rank seven}

There are four GPC for the three-electron system in
a rank-seven configuration $\wedge^3\H_7$, 
\begin{align}
D^1_{3,7} &= 2 - n_1 - n_2 - n_4 - n_7 \geq 0, \nonumber \\
D^2_{3,7} &= 2 - n_1 - n_2 - n_5 - n_6  \geq 0, \nonumber \\
D^3_{3,7} &= 2 - n_2 - n_3 - n_4 - n_5 \geq 0, \nonumber \\
D^4_{3,7} &= 2 - n_1 - n_3 - n_4 - n_6 \geq 0.
\label{eq:rankseven}
\end{align}

For lithium-like atoms, calculations~\cite{Sybilla,Mazziotti} had
shown that the first of these four inequalities is completely
saturated.

 \begin{figure}[!t]
 \centering 
 \includegraphics[width=6cm]{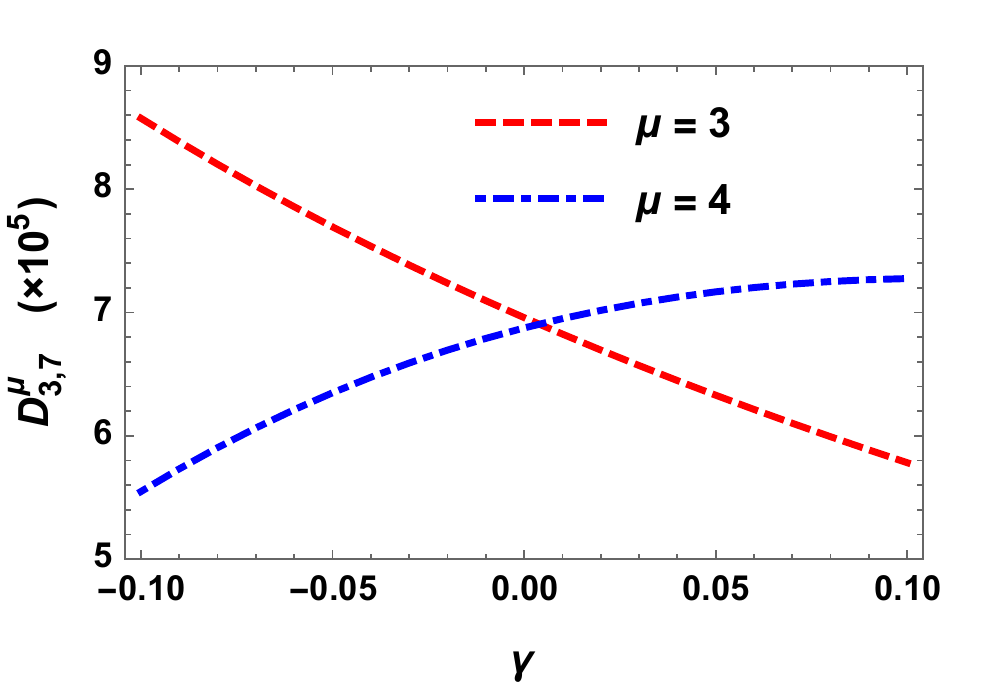}
 \caption{Third and fourth GPC for the ground-state of 
 the Hamiltonian $H(3,\ga)$ in a rank-seven approximation as a function 
 of $\ga \in [-0.1,0.1]$.}
 \label{figure5}
 \end{figure}
 
For the rank-seven approximation to the Hamiltonian~\eqref{eq:Hamiltonian},
we choose 
$$
\{\dl_1 \up, \dl_1 \dn, \psi \up,\dl_2 \up, \dl_2 \dn, \dl_3 \up, \dl_3\dn\}
$$
as the one-particle Hilbert space. Other types of configurations are
possible, too, but they lead to higher values for the ground-state
energy. There are 18 configurations in total, but only 14 of them
belong in the $j = \half$ representation. For any $\ga$, the
occupation numbers satisfy
\begin{align}
 n_1 + n_2 + n_4 + n_7 = 2 \word{and} n_3 + n_5 + n_6 = 1,
\label{eq:saturacion7}
\end{align}
implying that the first GPC of \eqref{eq:rankseven} is completely
saturated. The Hilbert space of this system then splits into the
direct product of two spin-orbital sectors $\wedge^3\,\H_7 = \H_3 \ox
\wedge^2\H_4$.

Also the following interesting system had been analyzed
in~\cite{Klyachko2}. The first excited state of beryllium, with spin
$(\textbf{S},\mathbf{S}_z)=(1,1)$, fills the lowest three shells $1s$,
$2s$ and $2p$. In a reasonable approximation, the first natural
orbital is completely occupied and the last two ones are empty (thus
$n_9 = n_{10} = 0$). The seven remaining natural orbitals are
organized in such a way that the first inequality
in~\eqref{eq:rankseven} is saturated, too.
 
 \begin{figure}[!t]
 \centering 
 \includegraphics[width=6cm]{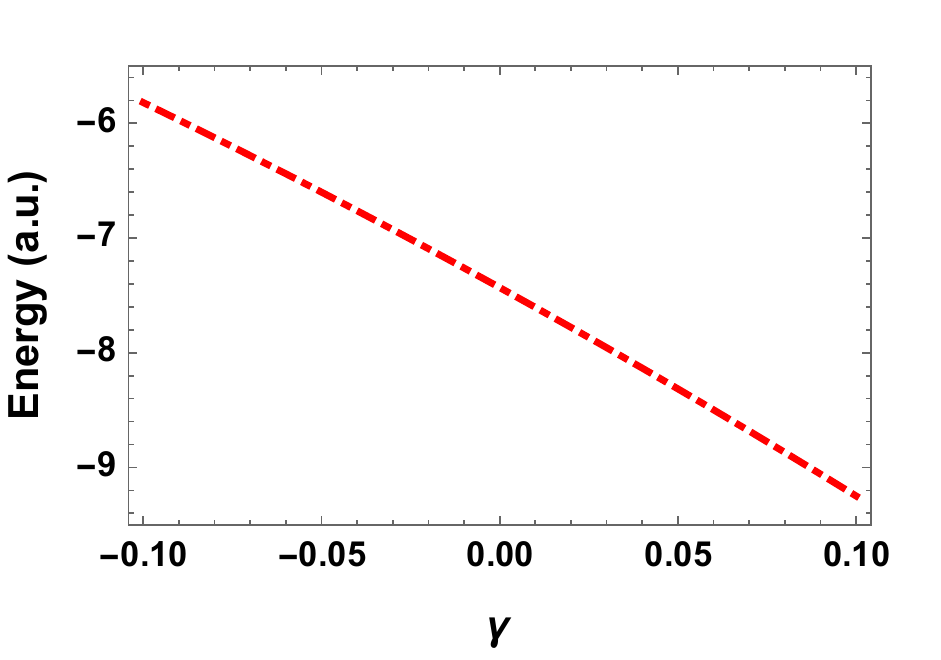}
 \caption{The ground-state energy for the spin-restricted rank-six 
 approximation to the Hamiltonian $H(3,\ga)$ given by \eqref{eq:Hamiltonian}
 as a function of $\ga \in [-0.1,0.1]$.}
 \label{figure6}
 \end{figure}
 
For lithium we found previously \cite{Sybilla} that the GPC could be
split into two groups differing in how close the equalities were
obeyed, i.e., one may talk about two scales of quasipinning. Here we
observe the same phenomenon. In fact, the value of the constraint
$D^2_{3,7}$ is always below $1.3717 \x 10^{-5}$, taking its maximum
for $\ga = 0.01$ (i.e., practically at the ``physical point''), as
indicated in Fig.~\ref{figure4}.

On the other hand, the remaining two GPC $D^3_{3,7}$ and $D^4_{3,7}$
take values around $7\x10^{-5}$. As shown in Fig.~\ref{figure5}, $D^3_{3,7}$
decreases when the value of $\ga$ grows, while the last one increases
when $\ga$ increases. Notice the crossover of two constraints also
close by $\ga=0$.

\subsection{On the energies}

Finally, Fig.~\ref{figure6} depicts the ground-state energy predicted by our model
for the spin-restricted version of the rank-six approximation as a
function of $\ga$. For $H(3,0)$ the ground-state energy is $-7.4311$
a.u. For rank seven and $\ga = 0$, the calculated energy for this
model is equal to $-7.4458$ a.u., lower than the Hartree--Fock energy
for lithium. Remarkably, the rank-eight approximation for this model
gives for the ground-state energy of lithium~$-7.4548$ a.u., which
represents more than $50\%$ of its correlation energy~\cite{Sybilla}.

\section{The molecular system $\mathrm{He}^+_2$}
\label{sec:molecular}

\begin{table*}[!htp]
 \centering      
{
\begin{tabular}{c c @{\hspace{10pt}} c @{\hspace{10pt}} c @{\hspace{10pt}} c @{\hspace{10pt}} c c c c c }  
Rank & Energy & $n_1$  & $ n_2$ & $n_3$ & $n_4 (10^{-3})$ & $n_5 (10^{-3})$ &
$n_6 (10^{-3})$   & $n_7 (10^{-3})$  & $n_8 (10^{-4})$ \\ [0.5ex] 
\hline\hline          
6 & $-4.9125$ & 0.9992 & 0.9949 & 0.9941 & 5.8086 & 5.0914 & 0.7172 &        -  & -    \\
7 & $-4.9194$ & 0.9973 & 0.9941 & 0.9915 & 7.1019 & 5.8950 & 2.5530 & 1.3220 & - \\
8 & $-4.9239$ & 0.9968 & 0.9932 & 0.9901 & 8.4888 & 6.8304 & 3.0819 & 1.3665 & 0.1178 \\
[1ex]   
\hline 
\end{tabular} }
\caption{Occupation numbers and energies for rank-six to rank-eight for He$_2^+$ in
its equilibrium geometry, employing 6-31G basis set.}
\label{table:litioON678} 
\end{table*}

In this section we study the behavior of the occupation numbers of
helium's molecular ion He$^+_2$. The goal is to explore the GPC along
the dissociation path of this three-electron system, whose symmetry is
lower than spherical, identifying those almost saturated. The
Hartree-Fock energy for this system is $-4.9$ a.u.~\cite{data}, with a
6-31G basis set. The equilibrium bond length is
$2.08$~a.u.~\cite{data}. The computed value for~the ground-state
energy is approximately $-4.99$~a.u.~\cite{exphe,data}. Therefore the
correlation energy is equal to $90$~mHa. We have approximated the
atomic orbitals by employing a 6-31G basis set \cite{STO}. We here
report our results for (rank six, seven and eight) CI approximations 
for this diatomic ion.

\subsection{GPC for the He$^+_2$ ground state}

For a dimer with atomic charges $Z$ the energy is 
given by the expression
\begin{align*}
&-\int \bigg( \frac12 \nabla^2_{\rr} + \sum_{\mu \in \{A,B\}}
\frac{Z}{|\rr -\RR_\mu|} \bigg) \, \rho_1(\xx,\xx')\bigg|_{\xx =\xx'}
\, d\xx
\\
&\qquad +\int \frac{\rho_2(\xx_1,\xx_2)}{|\rr_1 -\rr_2|} \, d\xx_1 \,
d\xx_2 + \frac{Z^2}{|\RR|}
\end{align*}
The two atoms are located at $\RR_A$ and $\RR_B$ and separated by
$\RR:= \RR_A - \RR_B$. The standard quantum-chemical notation $\xx :=
(\rr,\vs)$, with $\vs \in \{\up,\dn\}$ is employed. The molecular
orbitals are constructed as linear combinations of the atomic $1s$ and
$2s$ orbitals, which are in turn solutions of the Hartree-Fock
equations. In the rest of this subsection, standard notation for the
bonding (\textit{gerade}) and antibonding (\textit{ungerade})
molecular orbitals is used. The ground-state configuration of He$^+_2$
is classified as $^2\Sigma_u$ and the starting configuration is a
single Slater determinant, $\ket{1\sigma_g^{\up} 1\sigma_g^{\dn}
1\sigma^{\up}_u}$.

Table~\ref{table:litioON678} presents the results for the energy and
for the natural orbital occupancy numbers from rank-six up to
rank-eight approximations for the ground-state of He$_2^+$. The
rank-six approximation is obtained through a spin-compensated
configuration,
$$
\{1\sigma_g^\up, 1\sigma_g^\dn,1\sigma_u^\up,1\sigma_u^\dn,
2\sigma_g^\up,2\sigma_g^\dn\}.
$$
Higher-rank configurations are obtained by successively adding the
orbitals $\{2\sigma_u^\up, 2\sigma_u^\dn\}$.

\begin{table}[!b]
\centering    
{ 
\begin{tabular}{l @{\hspace{16pt}} r}  
\qquad Generalized Pauli conditions for $\wedge^3\H_8$  & $\x 10^{3}$  \\ [0.5ex] 
\hline\hline   
$0 \le \D^{1}_{3,8}  = 2 - (n_1 + n_2 + n_4 + n_7)$ &  0.0570 \\
$0 \le \D^{2}_{3,8}  = 2 - (n_1 + n_2 + n_5 + n_6)$ &  0 \\
$0 \le \D^{3}_{3,8}  = 2 - (n_2 + n_3 + n_4 + n_5)$ &  1.2712 \\
$0 \le \D^{4}_{3,8}  = 2 - (n_1 + n_3 + n_4 + n_6)$ &  1.4854 \\
$0 \le \D^{5}_{3,8}  = 1 - (n_1 + n_2 - n_3)$       &  0.0452 \\
$0 \le \D^{6}_{3,8}  = 1 - (n_2 + n_5 - n_7)$       &  1.2594 \\
$0 \le \D^{7}_{3,8}  = 1 - (n_1 + n_6 - n_7)$       &  1.4736 \\
$0 \le \D^{8}_{3,8}  = 1 - (n_2 + n_4 - n_6)$       &  1.3164 \\
$0 \le \D^{9}_{3,8}  = 1 - (n_1 + n_4 - n_5)$       &  1.5306 \\
$0 \le \D^{10}_{3,8} = 1 - (n_3 + n_4 - n_7)$      &   2.7449 \\
$0 \le \D^{11}_{3,8} = 1 - (n_1 + n_8)$               &  3.1772 \\
$0 \le \D^{12}_{3,8} = - (n_2 - n_3 - n_6 - n_7)$ &  1.3046 \\
$0 \le \D^{13}_{3,8} = - (n_4 - n_5 - n_6 - n_7)$ &  2.7901 \\
$0 \le \D^{14}_{3,8} = - (n_1 - n_3 - n_5 - n_7)$ &  1.5188 \\
$0 \le \D^{15}_{3,8} = 2 - (n_2 + n_3 + 2n_4 - n_5 - n_7 + n_8)$ &  7.7980 \\
$0 \le \D^{16}_{3,8} = 2 - (n_1 + n_3 + 2n_4 - n_5 - n_6 + n_8)$ &  5.9792 \\
$0 \le \D^{17}_{3,8} = 2 - (n_1 + 2n_2 - n_3 + n_4 - n_5 + n_8)$ &  5.0983 \\
$0 \le \D^{18}_{3,8} = 2 - (n_1 + 2n_2 - n_3 + n_5 - n_6 + n_8)$ &  3.0082 \\
$0 \le \D^{19}_{3,8} = - (n_1+ n_2 - 2n_3 - n_4 - n_5)$  &  5.4973 \\
[1ex]   
\hline    
\end{tabular} }
\caption{First 19 GPC for the system $\wedge^3\H_8$ and
numerical values for $\mathrm{He}^+_2$ at its equilibrium geometry.}
\label{table:M8} 
\end{table}

A number of findings can now be identified.

\begin{itemize}

\item For rank six, the spin-compensated configuration gives the
Borland--Dennis--Klyachko saturation condition $1 + n_3 = n_1 + n_2$.

\item For rank seven, we obtain the following values for the GPC:
\begin{align*}
D^1_{3,7} &= 2.42 \x 10^{-5}, \; D^2_{3,7} = 0, \\ 
D^3_{3,7} &= 1.24 \x 10^{-3}, \;
D^4_{3,7} = 1.39 \x 10^{-3}.
\end{align*}
The constraint due to spin has ``jumped'', with respect to the lithium 
series!

\item For this latter rank, two scales of quasipinning are clearly
identified. Compared with the lithium-like atom, the first level of
quasipinning is here more meaningful and probably more useful in order
to reduce the number of Slater determinants, since the 
distance to the ``Hartee-Fock'' point is here bigger, namely, 
$\xi = 1.06\x 10^{-2}$.

\end{itemize}

It is a fact of life that the number of GPC grows very rapidly with
rank. For rank eight there are \textit{31} inequalities
\cite{Alturulato}. They have been listed in a plain-text format
\cite{Data}. Of those, 19 constraints are given in
Table~\ref{table:M8}. The first four are equal to the Klyachko conditions for
$\wedge^3\H_7$. 

Several scales of quasipinning can be identified here, as well. Most
important is the \textbf{\textit{robustness}} of quasipinning. In
particular, the quantity $D^2_{3,8}$, found to be exactly zero in the
previous rank, remains in a saturated regime. The first and fifth
inequality belong to a strongly quasi-pinned regime, too. For the
remaining inequalities we have
\begin{align*}
&D^2_{3,8} \leq D^5_{3,8} \leq D^1_{3,8}\leq D^6_{3,8}\leq D^{3}_{3,8}
\\
&\leq D^{12}_{3,8} \leq D^8_{3,8}\leq D^{7}_{3,8} \leq D^{4}_{3,8}
\leq \cdots.
\end{align*}

\subsection{Occupation numbers and potential curves}

\begin{figure}[!t]
 \centering 
 \includegraphics[width=9cm]{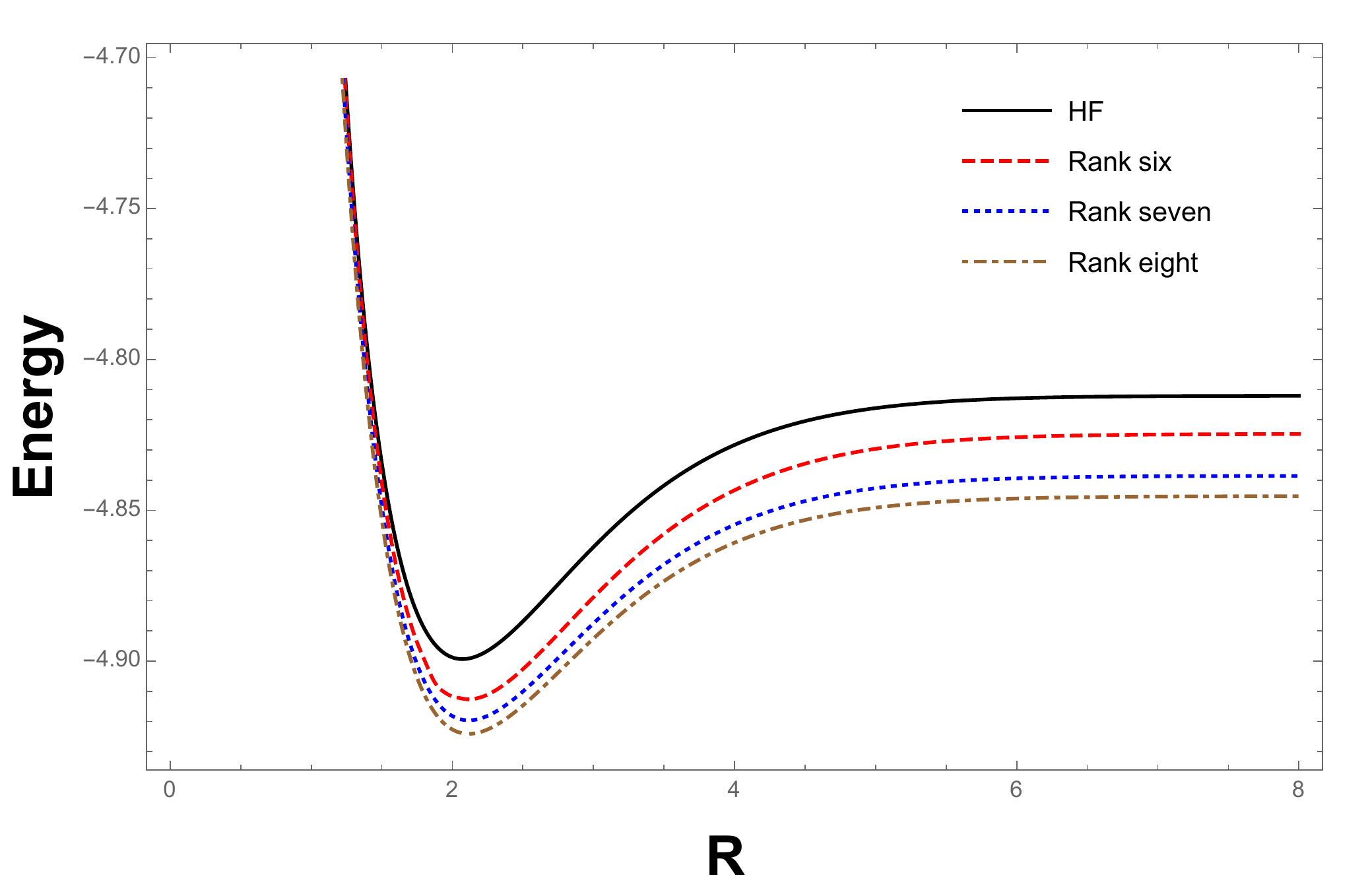}
 \caption{He$^+_2$ potential energy curves for the three ranks of CI approximation
 $\wedge^3\H_m$, $m\in\{3,6,7,8\}$ using 6-31G as a basis set.}
 \label{figurePE}
 \end{figure}

Potential energy curves for the three different ranks of the CI
approximation for He$^+_2$ are presented in Fig.~\ref{figurePE}. At the equilibrium
geometry, as also seen in Table~\ref{table:litioON678}, a larger rank
results in a lower ground-state energy. All approximations behave
similarly around the equilibrium distance. At large interatomic
distances, the value predicted by the rank-eight configuration is
$-4.845$ a.u.~which is to be compared with the total energy of the two
separated compounds (He and He$^+$): $-4.903$ a.u.~\cite{inonization}.

 \begin{figure}[!t] 
 \centering
  \includegraphics[width=7cm]{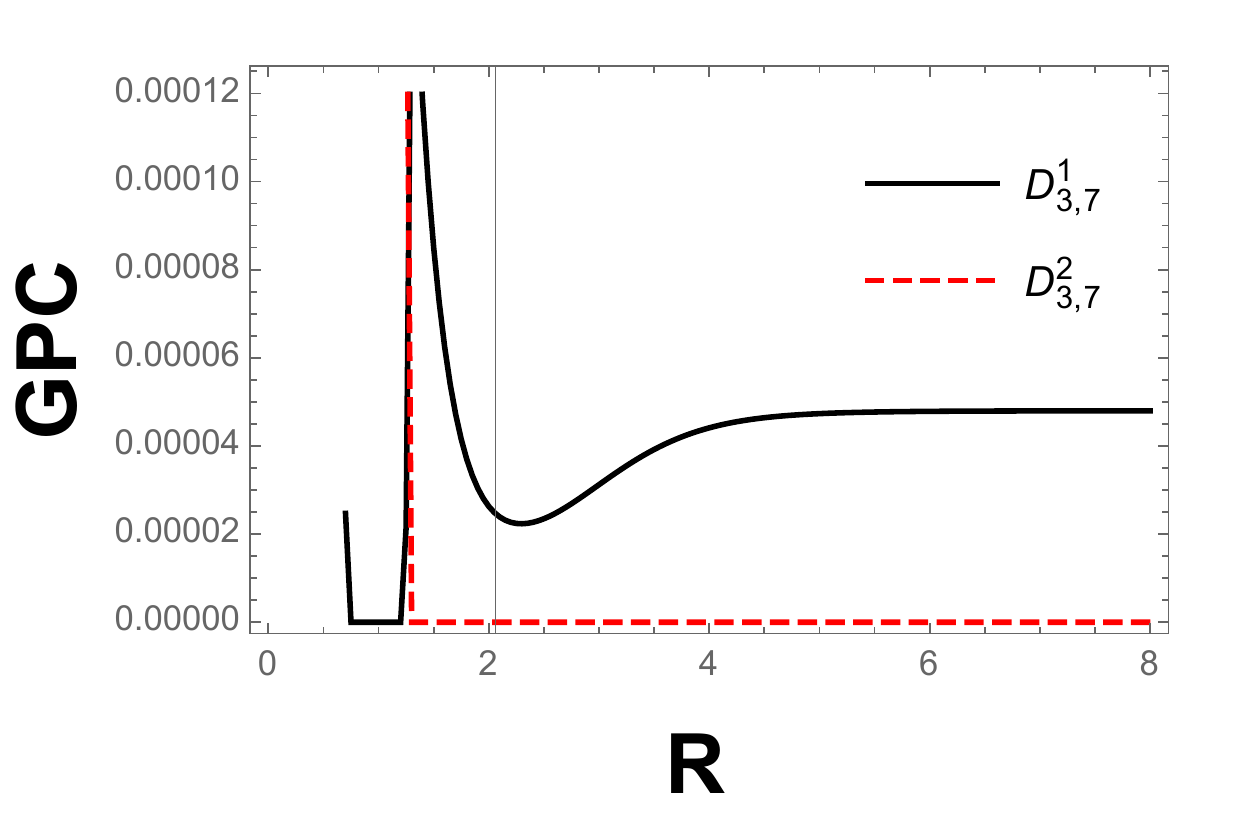} 
 \includegraphics[width=7cm]{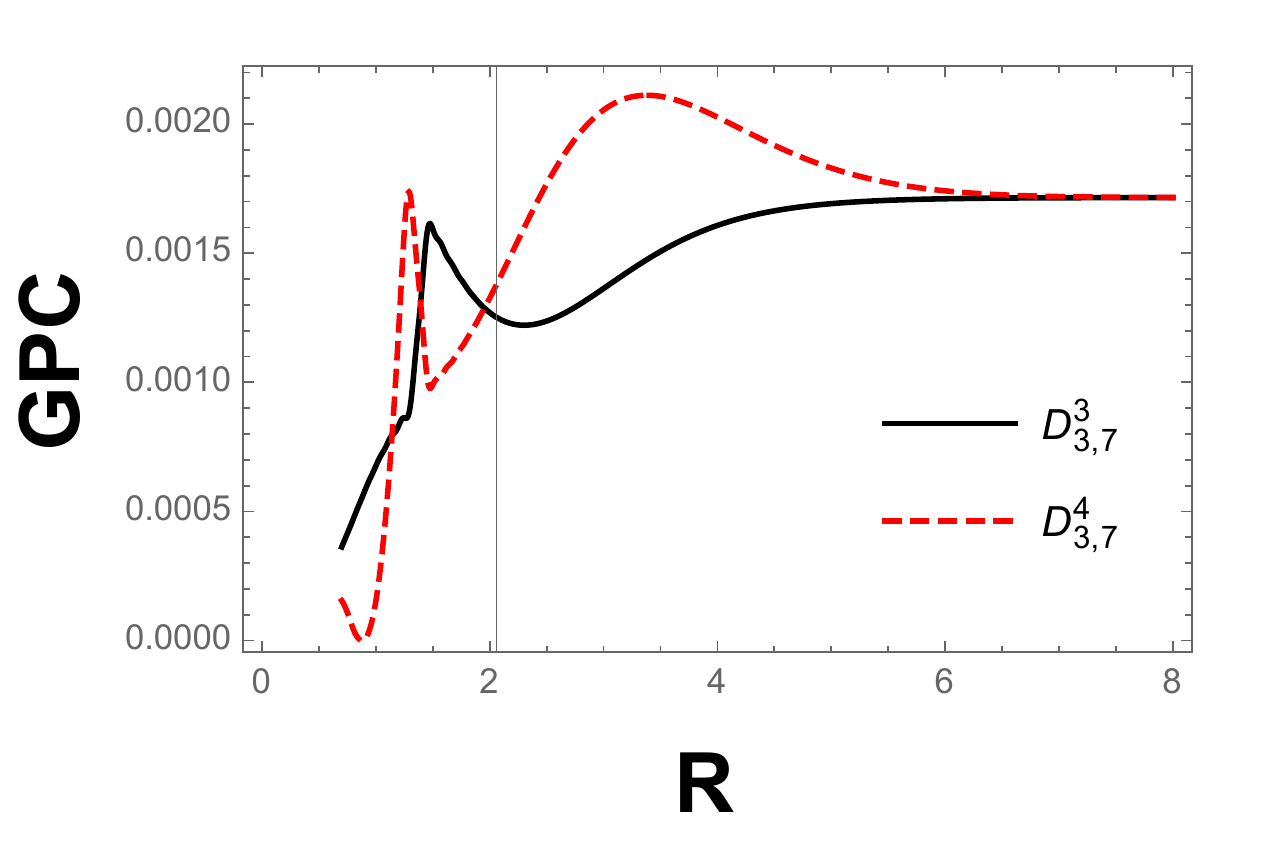}  
 \caption{Rank-seven GPC as functions of the interatomic 
 distance in atomic units. The vertical lines mark the equilibrium bond length.}
 \label{graf:figurerank7}
 \end{figure}

Fig.~\ref{graf:figurerank7} displays rank-seven GPC as functions of the interatomic
distance in atomic units. There are again two scales of quasipinning.
The first two GPC remain in a strongly pinned regime, since for those
$D_{3,7}^\mu$ is very close to 0. For those, we notice a sharp
crossover at lengths shorter than that of equilibrium. In fact, one of
them is \textit{always} completely saturated: in the region $R < 1.25$
a.u., i.e., $D^1_{3,7} = 0$ is a very good approximation, whereas for
$R > 1.25$ a.u. $D^2_{3,7} = 0$ is also very good. Unfortunately, we
do not have yet a good description for this apparent quenching of
degrees of freedom, which surely deserves further investigation.

 \begin{figure}[!t]
 \centering 
 \includegraphics[width=7cm]{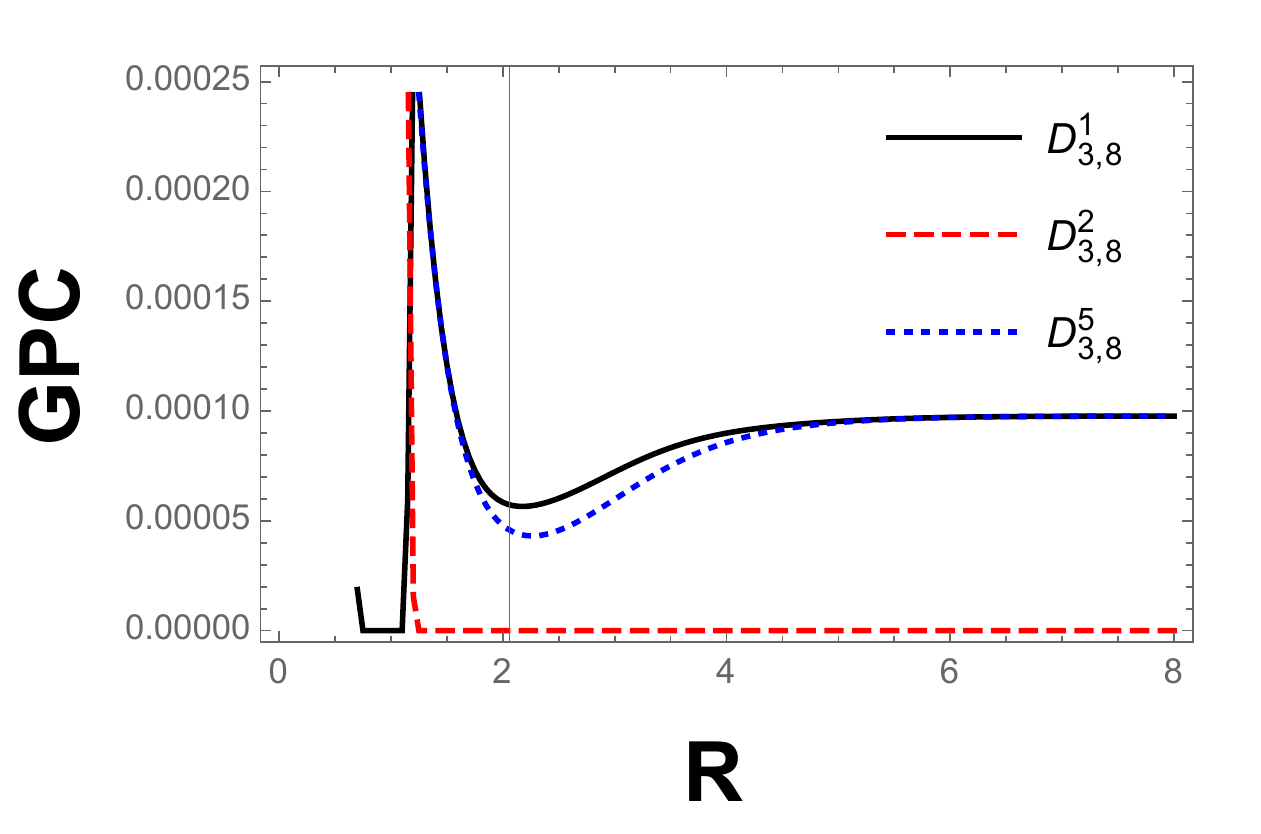}
 \includegraphics[width=7cm]{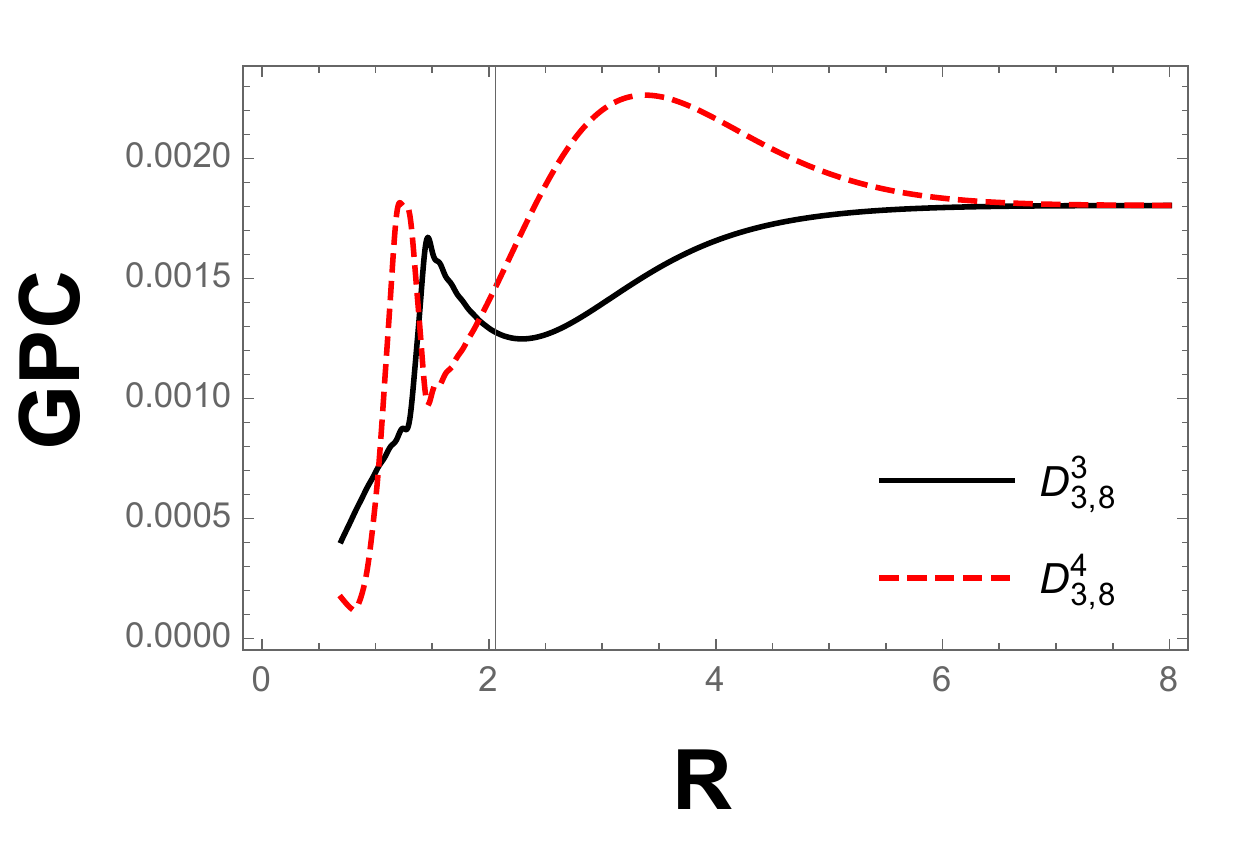}
 \includegraphics[width=7cm]{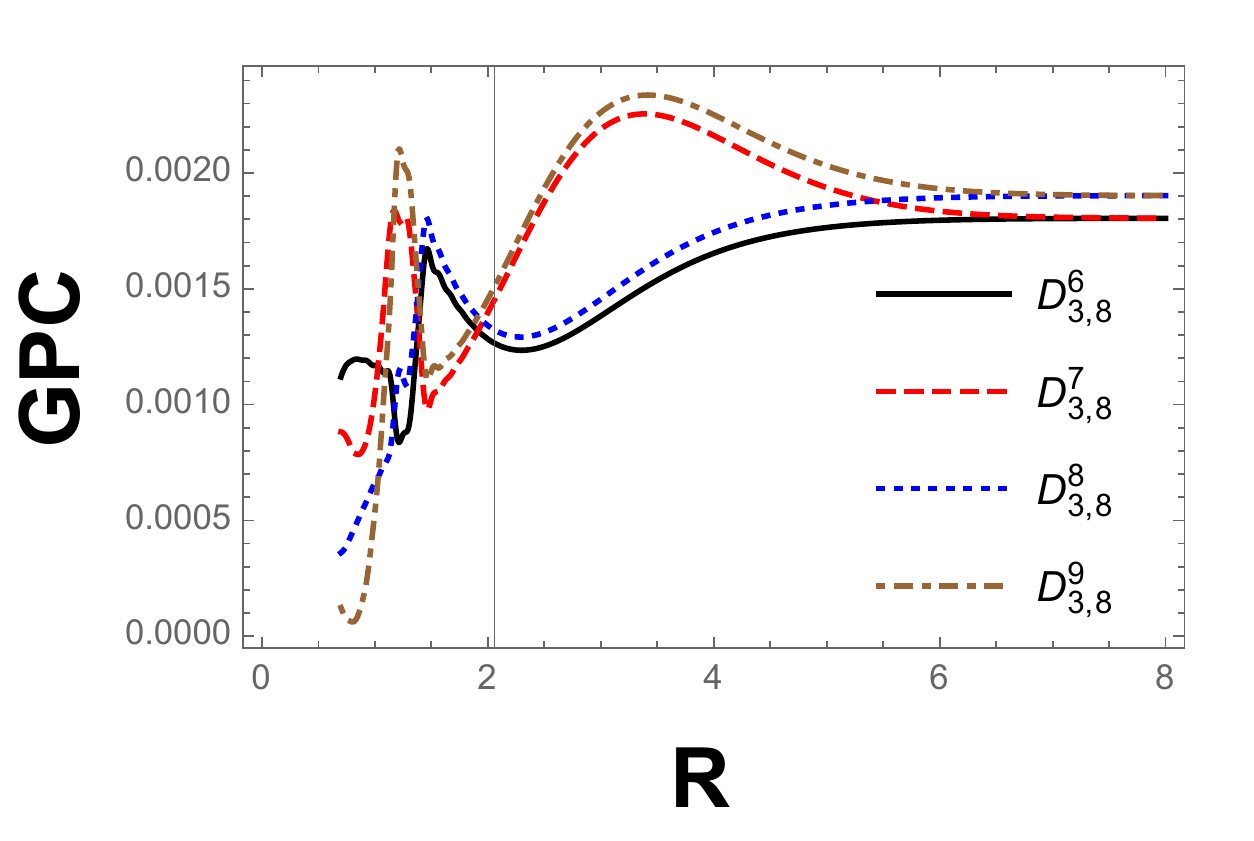}
 \includegraphics[width=7cm]{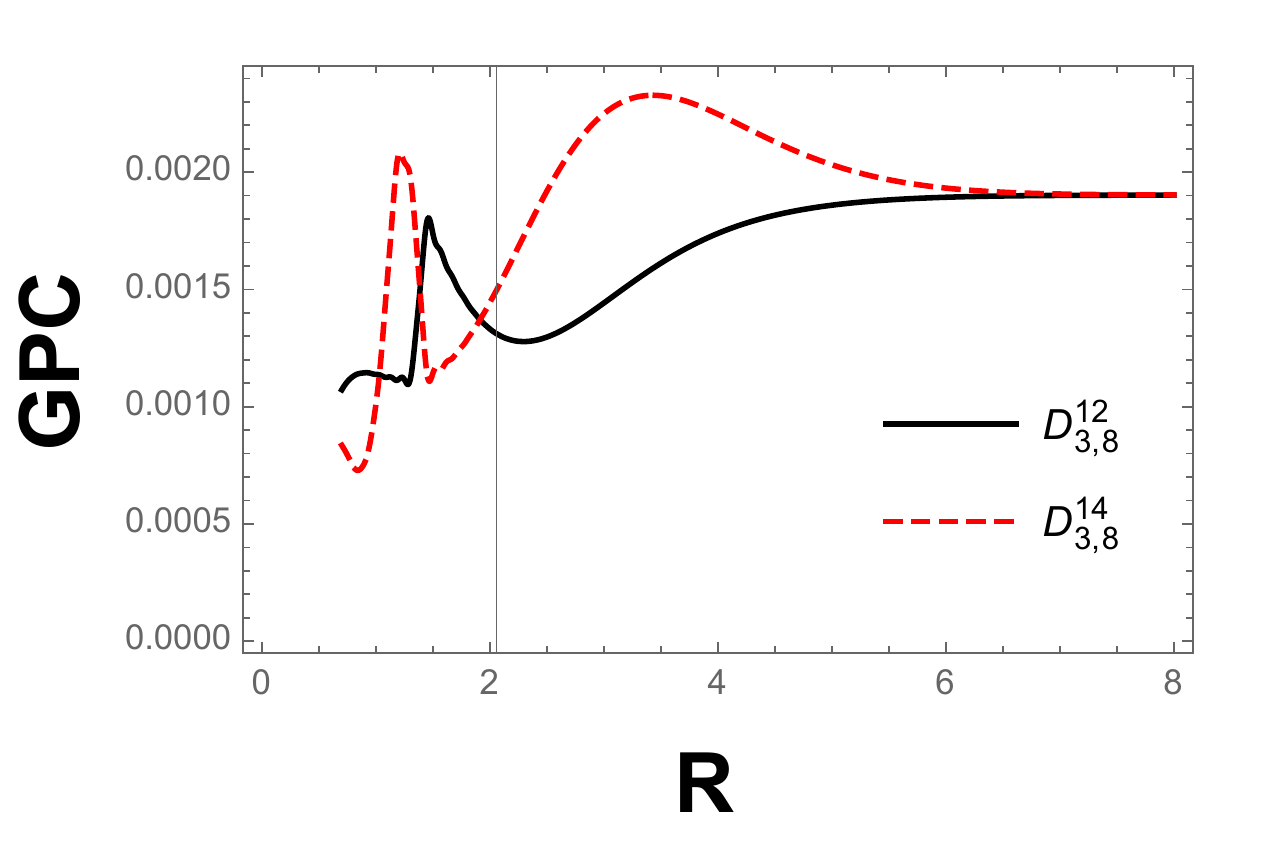}
 \caption{Rank-eight GPC as functions of the interatomic 
 distance. The vertical lines mark the equilibrium bond length.}
 \label{figurerank8}
 \end{figure}

For rank eight, several scales of quasipinning can be observed for
He$^+_2$. Our main result is again the robustness of quasipinning. In
particular, we observe that the quantities $D^1_{3,8}$ and
$D^2_{3,8}$, found to be exactly zero for some bond-length regime at
rank seven, remain in a strongly saturated regime, as shown in Fig.~9.
The Hilbert space of this system splits then into the direct product
of two spin-orbital sectors $\wedge^3\,\H_8 = \H_4 \ox \wedge^2\H_4$.
Also $D^5_{3,8}$ is found to be very close to 0.

To a second quasipinning regime belong the quantities 
$$
D^3_{3,8}, D^4_{3,8}, D^6_{3,8}, D^7_{3,8}, 
D^8_{3,8}, D^9_{3,8}, D^{12}_{3,8}, D^{14}_{3,8}.
$$
As seen in Fig.~\ref{figurerank8}, these GPC behave roughly in the
same way for increasing bond length. Their values tend asymptotically
to approximately the same value for large in\-te\-ratomic distances. 

Finally, a third quasipinning sector appears to be composed by
$D^{10}_{3,8}, D^{13}_{3,8}, D^{15}_{3,8}, D^{18}_{3,8},
D^{19}_{3,8}$.

\section{Quasipinning and excitations}
\label{excitations3}

From the seminal work by L\"owdin and Shull it is known that the
transformation to natural orbitals removes all single (S) excitations
of the wave function of two-electron systems \cite{LS}. For the
singlet state the general wave function can be written exactly as
$$
\ket{\Psi(\xx_1,\xx_2)} = \frac{1}{\sqrt{2}} (\up_1\dn_2 - \dn_1\up_2)
\sum^\infty_{i=1} c_i \ket{\al_i(\rr_1)\al_i(\rr_2)}.
$$
Again, we have used $\xx := (\rr, \vs)$ with $\vs$ being the spin
coordinates $\{\up,\dn\}$. A similar expression can be found for the
triplet state \cite{Laetitia}.

It is also remarkable that the wave function \eqref{eq:ranksix} does
not contain S or triple (T) excitations of the best single-determinant
state $\ket{0} := \ket{\al_1\al_2\al_3}$. The Slater determinants
$\ket{\al_1\al_4\al_5}$ and $\ket{\al_2\al_4\al_6}$ correspond to
double (D) excitations of this state.

Single excitations cannot be completely removed from the CI wave
function of general many-electron systems when written in terms of
natural orbitals. However, Mentel and coworkers \cite{Grillo} have
recently shown that writing the wave function in the basis of natural
orbitals leads to a sharp drop of the coefficients of Slater
determinants containing just S excitations. For the BH molecule, the
sum of squares of CI coefficients of singles falls from $1.5 \times
10^{-3}$ to $5.3 \times 10^{-6}$ when switching to the natural orbital
basis. In this section and the next we argue that this phenomenon is a
consequence from the near-saturation of some Klyachko selection rules on
the occupation numbers.

\subsection{Selection rule for excitations in $\wedge^3\H_6$}

This case has been just discussed. Even if the number of basis spin
orbitals pointing up is different of the number of the ones pointing
down, an eventual saturation of condition \eqref{eq:B&D} would lead to
the situation summarized in Table~\ref{table:SD36}. A double
excitation is also removed thereby.

\begin{table}[ht]
\centering      
{
\begin{tabular}{c c c c c c c c c c }  
Condition & $\ket{0}$ & S &  D & T & Total  \\ [0.5ex] 
\hline\hline           
CI                    & 1 & 3 & 3 & 1 & 8 \\
$\mathbf{D}^1_{3,6} \ket{\Psi}=0$   & 1 & 0 & 2 & 0 & 3  \\
[1ex]   
\hline    
\end{tabular} }
\caption{Number of Slater determinants in the total and force-pinned CI expansions 
of the wave function for the system $\wedge^3\H_6$.}
\label{table:SD36} 
\end{table}

\subsection{Selection rules for excitations in $\wedge^3\H_7$}

The four Klyachko inequalities for the three-electron system in
a rank-seven approximation $\wedge^3\H_7$ were given in
Eq.~\eqref{eq:rankseven}. The corresponding operators are
\begin{align*}
\mathbf{D}^1_{3,7} &= 2 - a^\7_1 a_1 - a^\7_2 a_2 -  a^\7_4 a_4 - a^\7_7 a_7,
\\
\mathbf{D}^2_{3,7} &= 2 - a^\7_1 a_1 - a^\7_2 a_2 -  a^\7_5 a_5 - a^\7_6 a_6, 
\\
\mathbf{D}^3_{3,7} &= 2 - a^\7_2 a_2 - a^\7_3 a_3 -  a^\7_4 a_4 - a^\7_5 a_5,
\\
\mathbf{D}^4_{3,7} &= 2 - a^\7_1 a_1 - a^\7_3 a_3 -  a^\7_4 a_4 - a^\7_6 a_6.
\end{align*}
As discussed above, for the lithium isoelectronic
series~\cite{Sybilla}, for the system described by the Hamiltonian of
Eq.~\eqref{eq:Hamiltonian} and for the first excited state of
beryllium in a rank-ten approximation~\cite{Klyachko2}, the first of
the four inequalities~\eqref{eq:rankseven} is completely saturated.
Accordingly, for all these systems, the exact wave function satisfies
the condition
$$
\mathbf{D}^1_{3,7}  \ket{\Psi}_{3,7}  = 0.
$$

This implies that in the natural orbital basis, every Slater
determinant is composed of three natural orbitals, two of them
belonging to the set $\{\al_1, \al_2,\al_4,\al_7\}$ and one belonging
to the set $\{\al_3,\al_5, \al_6\}$. Then, the system
$\wedge^3\H_7$ is reduced to $\H_3 \otimes \wedge^2\H_4$, with in
total eighteen of those Slater determinants.

Imposing as well saturation of the second inequality
of~\eqref{eq:rankseven}, i.e.,
$\mathbf{D}^2_{3,7}\mathbf{D}^1_{3,7}\ket{\tilde\Psi}_{3,7} =0$, the singles and
the triples are completely removed from the expression, as shown in
Table~\ref{table:SD37}. The corresponding wave function
$\ket{\tilde\Psi}_{3,7}$ is written in terms of the inicial configuration
$\ket{\al_1\al_2\al_3}$, plus the following eight D configurations:
\begin{align}
\ket{\al_1\al_4\al_5}, \ket{\al_2\al_4\al_6}, \ket{\al_1\al_5\al_7},
\ket{\al_2\al_5\al_7}, \nonumber \\
\ket{\al_1\al_4\al_6}, \ket{\al_2\al_4\al_5}, \ket{\al_1\al_6\al_7},
\ket{\al_2\al_6\al_7}.
\label{eq:conf7}
\end{align}

\begin{table}[ht]
 \centering      
{
\begin{tabular}{c c c c c c c c c c }  
Condition & $\ket{0}$ & S &  D & T & Total  \\ [0.5ex] 
\hline\hline           
$\mathbf{D}^1_{3,7}\ket{\Psi}_{3,7} =0$              & 1 & 6 & 9 & 2 & 18 \\
$\mathbf{D}^2_{3,7}\mathbf{D}^1_{3,7}\ket{\tilde\Psi}_{3,7} =0$   & 1 & 0 & 8 & 0 & 9  \\
[1ex]   
\hline    
\end{tabular} }
\caption{Number of Slater determinants in the total and force-pinned CI expansions
of the wave function for the system $\wedge^3\H_7$.}
\label{table:SD37} 
\end{table}

\subsection{Selection rules for excitations in $\wedge^3\H_8$}

The empirical evidence discussed earlier shows that the inequalities
for the following GPC are almost or completely saturated:
$$
D^1_{3,8}, D^2_{3,8}, D^5_{3,8}.
$$
Imposing the saturation of the second and fifth constraints, say,
the singles and the triples are removed completely, as shown in 
Table~\ref{table:SD38}. The corresponding wave function 
$\ket{\tilde\Psi}_{3,8}$ is written in terms of the 9 configurations of 
the pinned rank-seven wave function
\eqref{eq:conf7}, plus the configurations
\begin{align*}
\ket{\al_1\al_5\al_8}, \ket{\al_2\al_6\al_8}, \ket{\al_2\al_5\al_8},
\ket{\al_1\al_6\al_8}.  
\end{align*}

\begin{table}[!t]
 \centering      
{
\begin{tabular}{c c c c c c c c c c }  
Condition  & $\ket{0}$ & S &  D & T & Total  \\ [0.5ex] 
\hline\hline           
$\mathbf{D}^2_{3,8}\ket{\Psi}_{3,8}=0$  & 1 & 7 & 13 & 3 & 24 \\
$\mathbf{D}^5_{3,8}\mathbf{D}^2_{3,8}\ket{\tilde\Psi}_{3,8}=0$   & 1 & 0 & 12 & 0 & 13  \\
[1ex]   
\hline    
\end{tabular} }
\caption{Number of Slater determinants in the total and force-pinned CI expansions of 
the wave function for the system $\wedge^3\H_8$.}
\label{table:SD38} 
\end{table}

\subsection{He$^+_2$: electronic energy and pinning truncations}
\label{after truncating}

An idea behind quasipinning is to approximate the wave function
through a truncated expansion by using the selection ru\-les that
emerge after imposing pinning. Therefore, it is a relevant issue to
examine how the electronic energy is affected as the number of
configurations is reduced in the truncation. Here we ex\-plo\-re the
ground-state energy for the helium dimer He$^+_2$ for different pinned
wave functions, compared with the energy predicted by the CI expansion
\textit{within the same rank}. (It must be said beforehand that,
contrarily to the case of lithium-like systems, up to rank eight less
than 30\% of the absolute correlation energy is recovered. This is due
partly to a less than optimal choice of the basis set, partly to the
the difficulty to capture some aspects of correlation with such short
basis sets.)

Table~\ref{table:EnergySD37} contains the value of the correlation 
energy for (force-pinned and complete) wave functions for the rank-six 
up to -eight approximations for the ground state of He$^+_2$.

It is remarkable that the force-pinned wave function $\ket{\tilde\Psi}_{3,7}$
reconstructs $99.79$\%\, of the rank-seven correlation energy,
employing just 9 configurations. The CI rank-eight wave function
contains 24 Slater determinants belonging to the Hilbert space
$\wedge^2\H_4 \otimes \H_4$. The correlation energy is 24.64 mHa. The
pinned wave function $\ket{\tilde\Psi}_{3,8}$ reconstructs $99.51$\%\, of
this available correlation energy, employing 13 Slater determinants.

\begin{table}[!t]
 \centering      
{
\begin{tabular}{c @{\hspace{1em}} c @{\hspace{2em}} c}  
Wave function & $|E- E_\mathrm{HF}|$ \\ [0.5ex] 
\hline\hline           
$\ket{\Psi}_{3,6}$        & 13.22  \\
$\ket{\tilde\Psi}_{3,6}$        & 13.22  \\
$\ket{\tilde\Psi}_{3,7}$          & 20.12  \\
$\ket{\Psi}_{3,7}$          & 20.17 \\
$\ket{\tilde\Psi}_{3,8}$            & 24.56  \\
$\ket{\Psi}_{3,8}$                      & 24.64  \\
[1ex]   
\hline    
\end{tabular} }
\caption{Ground-state correlation energies predicted for the complete and force-pinned 
CI wave functions for He$^+_2$ in the rank-six up to rank-eight approximations. 
The values are given in mHa.}
\label{table:EnergySD37} 
\end{table}

\begin{figure}[!b] 
 \centering
  \includegraphics[width=8cm]{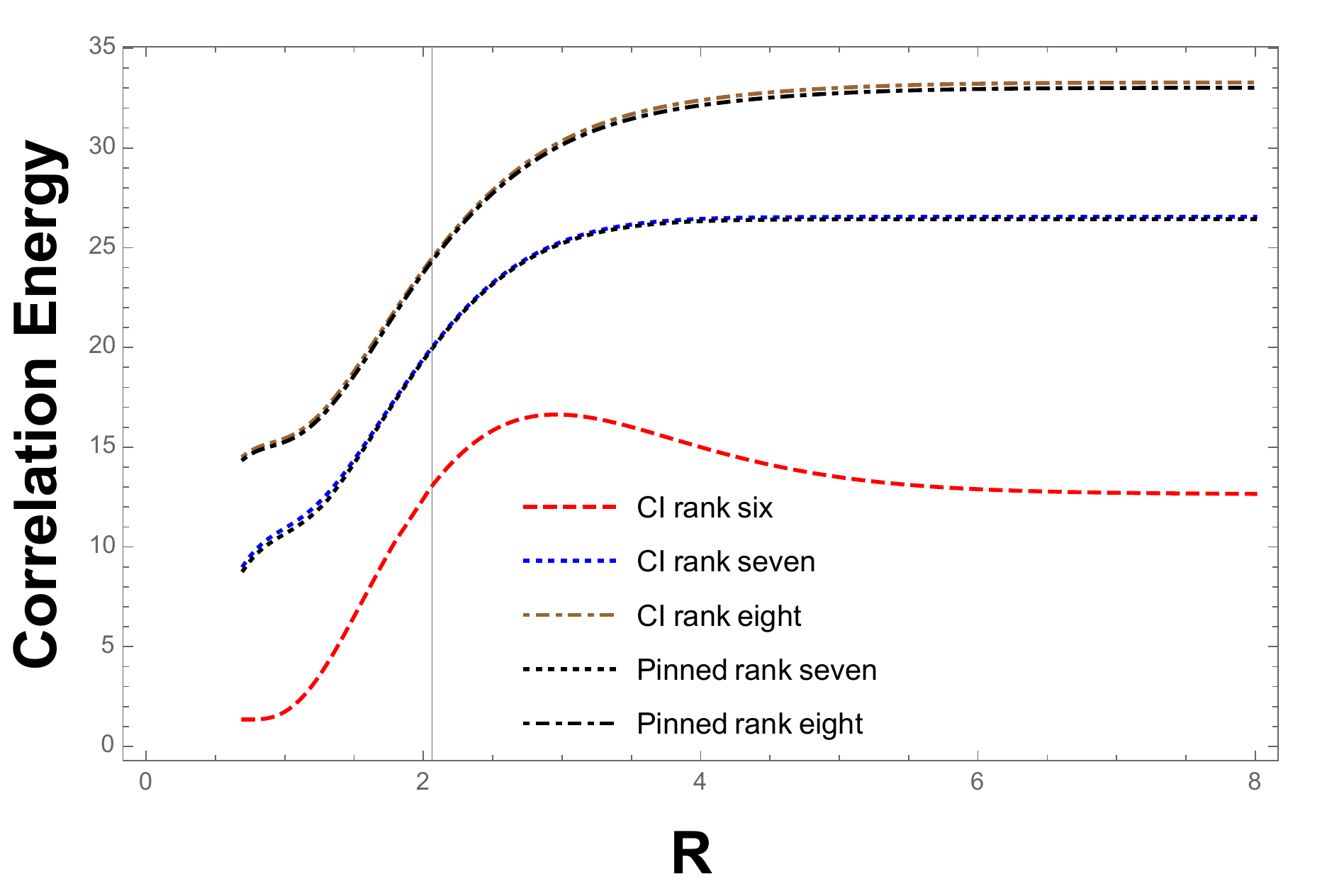} 
 \caption{Absolute value of the correlation energy  $|E - E_{HF}|$ 
 for CI expansion and pinned wave functions for ranks-six up to 
 rank-eight approximations, along the dissociation path for He$^+_2$.
 Pinned wave function $\ket{\tilde\Psi}_{3,6}$ is equivalent to the CI rank-six 
 expansion, so it is not included. The values are given in mHa.}
 \label{graf:CE}
 \end{figure}

Fig.~\ref{graf:CE} shows the absolute value of the correlation energy along the
dissociation path for CI rank-six up to rank-eight expansions
($\ket{\Psi}_{3,6}$, $\ket{\Psi}_{3,7}$, $\ket{\Psi}_{3,8}$) and for
the pinned wave functions $\ket{\tilde\Psi}_{3,7}$ and $\ket{\tilde\Psi}_{3,8}$.
It is also remarkable that the pinned rank-seven and rank-eight wave functions 
almost contain the complete correlation energy to the corresponding
 rank of approximation
along the complete path, demonstrating the negligible role of the
single and triple excitations. These results suggest that
in spite that saturation of one GPC reduces notably the number of
Slater determinants, remarkably good values for the correlation
energies are obtained.

\section{On four-electron systems}
\label{excitations4}

For the case of a four-electron system with a 8-dimensional
one-electron Hilbert space, $\wedge^4\H_8$, there are in total $14$
generalized Pauli conditions. Derived initially by
Klyachko \cite{Alturulato}, they read
\begin{align}
D^\mu_{4,8} &:= \sum^8_{i=1} \ka^\mu_i n_i \ge 0, \nonumber \\
D^{7+\mu}_{4,8} &:= 2 - \sum^8_{i=1} \ka^\mu_{9-i} n_i \ge 0,
\label{eq:48}
\end{align}
for $1 \le \mu \le 7$ and provided that $n_1 \le 1$.
The coefficients $\ka^\mu_i$ are given in Table \ref{table:four-eight}.

\begin{table}[b]
 \centering      
{
\begin{tabular}{c r r r r r r r r r}  
$\mu$  & $\ka^\mu_1$ & $\ka^\mu_2$ & $\ka^\mu_3$ & $\ka^\mu_4$ & $\ka^\mu_5$ 
& $\ka^\mu_6$ & $\ka^\mu_7$ & $\ka^\mu_8$ \\ [0.5ex] 
\hline\hline 
1 & $-1$ &      0 &      0 & 1 &      0 &      1 &      1 &      0  \\
2 & $-1$ &      0 &      0 & 1 &      1 &      0 &      0 &      1  \\
3 & $-1$ &      0 &      1 & 0 &      0 &      1 &      0 &      1  \\
4 & $-1$ &      1 &      0 & 0 &      0 &      0 &      1 &      1  \\
5 &      0 & $-1$ &      0 & 1 &      0 &      1 &      0 &      1  \\
6 &      0 &      0 & $-1$ & 1 &      0 &      0 &      1 &      1  \\
7 &      0 &      0 &      0 & 0 & $-1$ &      1 &      1 &      1  \\
[1ex]   
\hline
\end{tabular} }
\caption{Sets of coefficients for the generalized Pauli conditions of
\eqref{eq:48} for the system 
$\wedge^4\H_8$.}
\label{table:four-eight} 
\end{table}

For quantum states with an even number of fer\-mions, vanishing total
spin and ti\-me-reversal symmetry, Smith proved that a 1-RDM is
\textit{pure} $N$-representable if and only if all its eigenvalues are
doubly degenerated \cite{Smith}. Therefore, for these systems, the
occupation numbers obey
\begin{align}
n_{2i-1} = n_{2i}  \qquad \qquad i = 1, 2, \cdots.
\label{smith}
\end{align}
The double degeneracy of the occupation numbers forces the generalized
Pauli conditions for the system $\wedge^4\H_8$ to reduce to the
traditional Pauli exclusion principle \cite{Mazziotti}. Therefore, a
state will be pinned only if it is pinned to the traditional Pauli
conditions, which only occurs for a single-determinant wave function.
For instance,
\begin{align*}
D^1_{4,8} &:= -n_1 + n_4 + n_6 + n_7  = 2(1 - n_1),
\\
D^8_{4,8} &:= 2 - n_2 - n_3 - n_5 + n_8 = 2 n_8,
\\
D^{14}_{4,8} &:= 2 - n_1 - n_2 - n_3 + n_4 = 2 (1-n_1).
\end{align*}

Chakraborty and Mazziotti \cite{Mazziotti}~computed the occupation
numbers for the ground state of some four-electron molecules for rank
equal to twice the number of electrons, employing a STO-3G basis set.
In this range of approximation, the two energetically lowest orbitals
of LiH are completely occupied (therefore $D^1_{4,8} = 0$) and the
Shull--L\"ow\-din functional guarantees that doubly excited
determinants completely govern rank-eight CI calculations for this
molecule.

However, there are important effects of dynamical electron correlation
which involve the core electrons and the molecule cannot be
considered as a two-electron system. In fact, for higher ranks the two
biggest occupation numbers ($n_1 = n_2$) become smaller than 1. The
first (and the second as well) occupation number of BH is very close
to 1 and accordingly $D^1_{4,8}$ is quasipinned. For LiH and BeH$_2$,
the seventh occupation number is almost 0 and hence for these systems
$D^8_{4,8}$ is quasipinned.

In a spin-compensated description, the system $\wedge^4\H_8$ with
total spin component $\mathbf{S}_z$ equal to 1 contains 16
configurations, corresponding to $\wedge^3\H_4\otimes \H_4$. The CI
expansion only contains double or single excitations. 
In a spin-uncompensated description, the system $\wedge^4\H_8$ with
total spin component $\mathbf{S}_z$ equal to one would contain 30
configurations, corresponding to $\wedge^3\H_5\otimes \H_3$. 
Notice that if the~GPC
\begin{align}
D^{14}_{4,8} = 2 - n_1 - n_2 - n_3 + n_4 \ge 0 
\label{eq:ultima}
\end{align}
were completely saturated, the corresponding wave function is a member
of the 0-eigenspace of the operator:
\begin{align}
\mathbf{D}^{14}_{4,8} &= 2 - a^\7_1 a_1 - a^\7_2 a_2 -  a^\7_3 a_3 +  a^\7_4 a_4.
\end{align}
and, for both configurations, single and triple excitations are entirely suppressed. 
This is a non-trivial fact. See Tables \ref{table:SD48a} and \label{table:SD48b}.
Besides the initial configuration
$\ket{\al_1\al_2\al_3\al_4}$, the configurations present in the
expansion are just double excitations of this state which, in
addition, do not contain the fourth natural orbital $\al_4$.

\begin{table}[!t]
 \centering      
{
\begin{tabular}{c c c c c c c c c c }  
Condition & $\ket{0}$ & S &  D & T & Total  \\ [0.5ex] 
\hline\hline           
CI                                                   & 1 & 6 & 9 & 0  & 16 \\
$\mathbf{D}^{14}_{4,8}\ket{\Psi}=0$   & 1 & 0 & 9 & 0 & 10  \\
[1ex]   
\hline    
\end{tabular} }
\caption{Number of Slater determinants in the full and pinned CI expansions 
of the wave function for the spin-restricted system $\wedge^4\H_8$ with
$\mathbf{S}_z = 1$.}
\label{table:SD48a} 
\end{table}

In general, for the system $\wedge^N\H_m$, the condition
$$
(N-2) + n_N = n_1 + \cdots + n_{N-1}
$$ 
has as consequence that only double excitations become the relevant
configurations in a CI expansion \cite{Kassandra}. Moreover, the only
configuration containing the orbital $\al_N$ is $\ket{\al_1\al_2
\cdots\al_N}$.

\begin{table}[!t]
 \centering      
{
\begin{tabular}{c c c c c c c c c c }  
Condition & $\ket{0}$ & S &  D & T & Total  \\ [0.5ex] 
\hline\hline           
CI                                                   & 1 & 8 & 16 & 5  & 30 \\
$\mathbf{D}^{14}_{4,8}\ket{\Psi}=0$   & 1 & 0 & 11 & 0 & 12  \\
[1ex]   
\hline    
\end{tabular} }
\caption{Number of Slater determinants in the full and pinned CI expansions 
of the wave function for the spin-unrestricted system $\wedge^4\H_8$ with
$\mathbf{S}_z = 1$.}
\label{table:SD48b} 
\end{table}

For the larger system $\wedge^4\H_{10}$, the occupation numbers are
bounded by 121 constraints~\cite{Alturulato, Data}. We postpone their 
study.

\section{Conclusion}
\label{conclusion}

The recent solution of the pure $N$-representability
problem, due to Klyachko, promises to generate a wide set of
conditions (the GPC) on the natural occupation numbers for fermionic
systems. The Klyachko algorithm does indeed produce sets of linear
inequalities with integer coefficients for those numbers. The
derivation of these inequalities, and of their consequences, is still
a work in progress.

For reasons that nobody has been quite able to fathom yet, some of
these inequalities appear to be nearly saturated, in a far from random
way ---this is the quasipinning phenomenon. A research program is born
around these facts.

By means both of theoretical and numerical results, in this paper we
have continued to explore the nature of pinning and quasipinning in
some atomic and molecular models (mainly perturbed lithium with broken
spherical symmetry and the dimer ion He$^+_2$), for several finite
rank approximations whose GPC are known.

We sum up our opinions on the outcomes of that program, so far.

\begin{itemize}

\item{} Saturation of some of the GPC leads to strong selection rules
for identifying the most (in)effective configurations in CI
expansions. In simple cases, this gives means for reducing the number
of Slater determinants in the CI picture and therefore reducing
computational requirements \cite{ETH,Sybilla,Klyachko2}. In general,
it does provide insights in the structure of the wave function, which
brute force methods are unable to.

\item{} However, it is unlikely that Klyachko paradigm be relevant for
computational quantum chemistry, at least in the short run. The main
problem is the dramatic increase of the number of GPC with the rank of
the spin orbital systems introduced in the calculations.

\item{} The robustness of the almost saturation of a particular type
of constraint conspires to ``explain'' why double excitations govern
CI calculations of electron correlation, when using natural orbitals.

\item{} A natural question is whether the exact ``L\"owdin--Shull''
formula~\eqref{eq:yupi!} for three-electron systems can be generalized
to higher rank. The answer is a qualified, approximate yes, the price
to pay being to invoke a second type of constraint less strongly
quasi-pinned that the one referred to in the previous point. We
refrain form going into the details here.

\item{} A very promising avenue of research is to use the GPC to
improve on the 1-RDM theory. There are now in the literature quite a
few physically motivated density matrix functionals, built from the
knowledge of the natural orbitals and occupation numbers, which can be
traced back to the one proposed by M\"uller thirty years ago
\cite{Muller}; they have mostly amounted to figure out
\textit{Ans\"atze} for reasonable two-body reduced density matrices,
failing to date to fulfill a physical requirement or another \cite{Marmulla}. 
The
approach discussed in this paper suggests to construct 1-RDM by
restricting the minimization set to the subset of GPC-honest systems.
A promising start in this direction is~\cite{Halle}.

\end{itemize}



\begin{acknowledgements}
The authors are most grateful to E.~R.~Davidson,
J.~M.~Gracia-Bond\'ia, D.~Gross, R.~Herrero-Hahn, S.~Kohaut,
D.~A.~Mazziotti, C.~Schilling, J.~C.~V\'arilly and M.~Walter for
helpful comments and illuminating discussions. We thank N. Louis for
IT support. CLBR was supported by Colombian Department of Sciences and
Technology (Colciencias). He very much appreciates the warm atmosphere
of the Physikalische und Theoretische Chemie group at Saarlandes
Universit\"at.
\end{acknowledgements}



\begin{thebibliography}{30}
\bibitem{Ccomment}
C. Schilling pointed to us at once that there is no ambiguity of signs
in display~\eqref{eq:yupi!}.
\end{thebibliography}
\end{document}